\documentclass[11pt,cls,onecolumn]{IEEEtran}

\usepackage{subfigure,cite,graphicx,amsmath,amssymb,eufrak,mathrsfs,epsfig,dblfloatfix}
\usepackage{amsmath,amssymb,mathrsfs,graphicx}
\usepackage{psfrag}
\usepackage{multirow}

 \DeclareMathOperator\E{\sf E}
\let\P\relax
\DeclareMathOperator\P{\sf P}

\newtheorem{theorem}{Theorem}
\newtheorem{example}{Example}

\newtheorem{lemma}{Lemma}

\begin{document}
\title{Approximate Ergodic Capacity of a Class of Fading $2$-user $2$-hop Networks}

\author{Sang-Woon Jeon\IEEEmembership{, Member, IEEE}, Chien-Yi Wang\IEEEmembership{, Student Member, IEEE}, and\\ Michael Gastpar\IEEEmembership{, Member, IEEE}
\thanks{This work has been supported in part by the European ERC
Starting Grant 259530-ComCom.}
\thanks{The material in this paper was presented in part at the Information Theory and Applications Workshop (ITA), San Diego, CA, February 2012 and the IEEE International Symposium on Information Theory (ISIT), Boston, MA, July 2012.}
\thanks{S.-W.~Jeon, C.-Y. Wang, and M. Gastpar are with the School of Computer and Communication Sciences,
Ecole Polytechnique F{\'e}d{\'e}rale de Lausanne (EPFL), Lausanne,
Switzerland (e-mail: \{sangwoon.jeon, chien-yi.wang, michael.gastpar\}@epfl.ch).}%
\thanks{M. Gastpar is also with the Department of Electrical Engineering and Computer Sciences, University of California, Berkeley, CA, USA.}
}
\maketitle

\IEEEpeerreviewmaketitle

\begin{abstract}
We consider a fading AWGN 2-user 2-hop network where the channel
coefficients are independent and identically distributed (i.i.d.) drawn from a continuous distribution and
vary over time.
For a broad class of channel distributions, we characterize the
ergodic sum capacity to within a constant number of bits/sec/Hz,
independent of signal-to-noise ratio.
The achievability follows from the analysis of an interference
neutralization scheme where the relays are partitioned into $M$ pairs,
and interference is neutralized separately by each pair of relays.
When $M=1$, the proposed ergodic interference neutralization
characterizes the ergodic sum capacity to within $4$ bits/sec/Hz for
i.i.d. uniform phase fading
and approximately $4.7$ bits/sec/Hz for i.i.d. Rayleigh fading.
We further show that this gap can be tightened to $4\log \pi-4$ bits/sec/Hz (approximately $2.6$) for i.i.d. uniform phase fading and $4-4\log( \frac{3\pi}{8})$ bits/sec/Hz (approximately
$3.1$) for i.i.d. Rayleigh fading in the limit of large $M$.\footnote{Throughout the paper, $\log(\cdot)$ denotes the logarithm of base two.}
\end{abstract}


\section{Introduction}
In recent years, there has been significant progress towards understanding fundamentals of multi-source single-hop networks \cite{Etkin:08,Viveck1:08,Nazer11:09}.\footnote{Unless otherwise specified, we assume Gaussian networks throughout the paper.}
Following up on these successes for single-hop networks, more recent and emerging work has considered \emph{multi-source multi-hop networks} \cite{Mohajer:11,Tiangao:12,Jeon2:11,Wang:11,Shomorony:11}.
For multi-source multi-hop networks, interference can be cancelled by aligning multiple paths through the network, a technique referred to as \emph{interference neutralization}.
Proper exploitation of such interference neutralization is the key for an approximate capacity \cite{Mohajer:11} and the optimal degrees of freedom (DoF) characterization \cite{Tiangao:12,Jeon2:11,Wang:11,Shomorony:11}.
Recently, for $2$-user $2$-hop networks, interference neutralization combining with symbol extension was used to show that two relays suffice to achieve the optimal DoF \cite{Tiangao:12}.
In spite of recent progress in this area, the best known capacity characterization for fully connected $2$-user $2$-hop networks is to within $o(\log(\mbox{SNR}))$ bits/sec/Hz \cite{Tiangao:12}, which can be arbitrarily large as the signal-to-noise ratio (SNR) increases.

The aim of this paper is to \emph{tighten the capacity gap of $2$-user $2$-hop networks  to within a constant number of bits/sec/Hz, independent of SNR}.
Our achievability is based on \emph{ergodic interference neutralization} \cite{Jeon2:11}, which is similar to ergodic interference alignment \cite{Nazer11:09} applied to multi-source single-hop networks.
Suppose that the sources transmit their signals at time $t$ through the first-hop channel matrix $\mathbf{H}[t]$.
Then the relays amplify and forward their received signals with an appropriate delay $\tau$ through the second-hop channel matrix  $\mathbf{G}[t+\tau]$ such that $\mathbf{G}[t+\tau] \mathbf{H}[t]$ becomes an approximately diagonal matrix with non-zero diagonal elements.
This approach can completely neutralize interference in the finite SNR regime.

Assuming independent and identically distributed (i.i.d.) channel coefficients, the proposed ergodic interference neutralization characterizes the ergodic sum capacity to within 
a constant number of bits/sec/Hz for a broad class of channel distributions.
For instance, when the number of relays $L$ is equal to two, it achieves the ergodic sum capacity to within $4$ bits/sec/Hz for uniform phase fading and  approximately $4.7$ bits/sec/Hz for Rayleigh fading.
As $L$ increases, we narrow the corresponding gap in our analysis.
Specifically, this gap is given as $4\log \pi-4$ bits/sec/Hz (approximately $2.6$) for i.i.d. uniform phase fading and $4-4\log( \frac{3\pi}{8})$ bits/sec/Hz (approximately
$3.1$) for i.i.d. Rayleigh fading in the limit of large $L$.
We also notice that a similar analysis is applicable for the $K$-user interference channel and show that ergodic interference alignment in \cite{Nazer11:09} characterizes the ergodic sum capacity assuming that all sources employ uniform power allocation across time to within $(\frac{1}{2}\log 6)K$ bits/sec/Hz (approximately $1.3K$) for i.i.d. Rayleigh fading.
Table \ref{table:summary} summarizes the new approximate ergodic capacity results of this paper and the existing DoF and approximate capacity results.

\begin{table}[t!]
\scriptsize
\caption{New approximate capacity results and the existing DoF and approximate capacity results.}
\begin{center}
\begin{tabular}{c|c|c|c|c} \label{table:summary}
&\multirow{1}{*}{$K$-user} &$2$-user $2$-hop
network&$K$-user $2$-hop network& $K$-user
$K$-hop network\\
&interference channel&with $2$ relays&with $L$ relays& with  $K$ relays at each layer \\
\hline\hline
\multirow{3}{*}{DoF}&\multirow{3}{*}{$\frac{K}{2}$
\cite{Viveck1:08}}&\multirow{3}{*}{$2$\cite{Tiangao:12}}& Generally unknown& \multirow{2}{*}{Generally unknown}  \\
&&&$K$ if $L\geq K(K-1)+1$\cite{Rankov:07} &\multirow{2}{*}{ $K$ for  isotropic fading\cite{Jeon2:11}}\\
&&& $L$ if $K\to\infty$ \cite{Viveck2:09}\\
\hline\hline
\multirow{1}{*}{Ergodic capacity}&\multirow{3}{*}{Exact capacity\cite{Nazer:09}}&\multirow{3}{*}{$4$ bits/sec/Hz gap}&Generally unknown \\
\multirow{1}{*}{for uniform}&&&$2.6$ bits/sec/Hz gap &\multirow{1}{*}{Unknown}   \\
\multirow{1}{*}{phase fading}&\multirow{1}{*}{}&\multirow{1}{*}{}& if $K=2$ and  $L\to\infty$&\\
\hline
\multirow{2}{*}{Ergodic capacity}&\multirow{3}{*}{$1.3K$ bits/sec/Hz gap}&\multirow{3}{*}{$4.7$ bits/sec/Hz gap}&Generally unknown \\
\multirow{2}{*}{for Rayleigh fading}&&&$3.1$ bits/sec/Hz gap &\multirow{1}{*}{Unknown} \\
&\multirow{1}{*}{}& \multirow{1}{*}{}& if $K=2$
and $L\to\infty$ & \\
\hline
\end{tabular}
\end{center}
\label{default}
\end{table}%

\subsection{Related Work}

\subsubsection{Degrees of freedom}
In seminal work \cite{Viveck1:08}, interference alignment has been proposed to achieve the optimal DoF of the $K$-user interference channel with time-varying channel coefficients.
The concept of this signal space alignment has been successfully adapted to various network environments, e.g., see \cite{Maddah-Ali:08,Viveck2:09,Tiangao:10,Suh:11, Suh:08,Viveck1:09,Annapureddy:11,Ke:12} and the references therein.
It was shown in \cite{Motahari:09, Motahari2:09} that interference alignment can also be attained on fixed (not time-varying) interference channels.  

In spite of recent achievements on interference channels or multi-source single-hop networks,
understanding of multi-source multi-hop networks is still in progress.
The work \cite{Tiangao:12} has exploited interference alignment to neutralize interference at final destinations, which is referred to as aligned interference neutralization, and showed that the optimal $2$ DoF is achievable for $2$-user $2$-hop networks with $2$ relays.
This result has been recently generalized to two unicast networks \cite{Wang:11,Shomorony:11}.
For more than two unicast, the optimal DoF is in general unknown except for a certain class of networks.
For the $K$-user $2$-hop network with $L$ relays, interference can be completely neutralized if $L\geq K(K-1)+1$ \cite{Rankov:07}.
Similar concept of ergodic interference alignment has been proposed for interference neutralization in \cite{Jeon2:11} showing that ergodic interference neutralization achieves the optimal DoF of $K$-user $K$-hop isotropic fading networks with $K$ relays in each layer.

\subsubsection{Beyond degrees of freedom}
The DoF discussed previously is a fundamental
metric of multi-source networks especially for high SNR, which characterizes capacity to within $o(\log \mbox{SNR})$ bits/sec/Hz.
Depending on the operational regime, however, the gap of $o(\log \mbox{SNR})$
bits/sec/Hz in practice can be significant and achieving the optimal
DoF may not be enough.
For the $2$-user interference channel, for instance, time-sharing
between the two users can also achieve the optimal one DoF.
On the other hand, a simple Han--Kobayashi scheme can tighten the gap to within
one bit/sec/Hz \cite{Etkin:08}, which provides an arbitrarily larger
rate compared with the time-sharing for a certain operational regime and
channel parameters.
Consequently, several works have recently established tighter bounds
on the gap from capacity \cite{Bresler:10,Nam:10,Avestimehr:11,Lim:11,Mohajer:11,Niesen2:11,Ordentlich:12} to provide a universal performance guarantee, independent of SNR
and channel parameters.

A similar flavor of such bounds on the gap from capacity concerns time-varying channel models. 
The recently proposed ergodic interference alignment in \cite{Nazer11:09} makes interference aligned in the finite SNR regime and, as a result, provides significant rate improvement compared with the conventional time-sharing strategy in the finite SNR regime.
Ergodic interference alignment was shown to achieve the ergodic sum capacity of the $K$-user interference channel for i.i.d. uniform phase fading \cite{Nazer11:09}.
For the $K$-user finite field interference channel (with time-varying channel coefficients), the idea of ergodic interference alignment was independently proposed by Nazer \emph{et al.} \cite{Nazer:09} and Jeon and Chung \cite{JeonITA:09} in two slightly different versions.
In \cite{Niesen:11}, ergodic channel pairing was applied to tighten the gap from the ergodic capacity for fading multihop networks showing a gap depending only on the number of nodes in a layer, instead of the total number of nodes in a network.

\subsection{Paper Organization}
The rest of the paper is organized as follows.
In Section \ref{sec:setup}, we  introduce the fading $2$-user $2$-hop network model considered in this paper and formally define its ergodic sum capacity. 
In Section \ref{sec:main_results}, we first state the main results of the paper, approximate ergodic sum capacities of fading $2$-user $2$-hop networks.
In Section \ref{sec:ergodic_in}, we explain ergodic interference neutralization and its achievable rate. 
In Section \ref{sec:appro_cap}, we prove the approximate ergodic sum capacity results in Section \ref{sec:main_results} based on the achievability in Section \ref{sec:ergodic_in}.
Finally, we conclude in Section \ref{sec:conclusion} and refer some technical proofs to the appendices.

\begin{table}[t!]
\caption{Summary of notation} \label{Table:symbols}
\vspace{-0.1in}
\begin{equation*}
    \begin{array}{|c|c|}
    \hline
    \mathbf{A}^{T} (\mbox{ or }\mathbf{a}^{T})& \mbox{Transpose of }\mathbf{A}(\mbox{ or } \mathbf{a})\\
    \hline
    \mathbf{A}^{\dagger} (\mbox{ or }\mathbf{a}^{\dagger})  & \mbox{Conjugate transpose of }\mathbf{A}(\mbox{ or }    \mathbf{a})\\
    \hline
    \det(\mathbf{A}) & \mbox{ Determinant of $\mathbf{A}$ }\\
    \hline
    \mathbf{I} & \mbox{ Identity matrix }\\
    \hline
    \jmath & \sqrt{-1}\\
    \hline
    \operatorname{re}(a)(\mbox{ or }\operatorname{im}(a))& \mbox{Real (or imaginary) part of } a\\
    \hline
    |a| & \mbox{Absolute value of } a\\
    \hline
    a^* & \mbox{Complex conjugate of } a\\
    \hline
    \lfloor a\rfloor & \mbox{Floor of } a{~}(\lfloor a\rfloor=\max\{x\in\mathbb{Z}|x\leq a\})\\
    \hline
    \operatorname{card}(\mathcal{A})&\mbox{ Cardinality of } \mathcal{A}\\
    \hline
    \mathcal{CN}(\mu,\sigma^2) &\mbox{ Circularly symmetric complex Gaussian distribution with mean $\mu$ and variance $\sigma^2$ } \\
    \hline
    \end{array}
\end{equation*}
\end{table}

\section{Problem Formulation} \label{sec:setup}
In this section, we explain our network model and define its sum capacity.
Throughout the paper, we will use $\mathbf{A}$, $\mathbf{a}$, and $\mathcal{A}$ to denote a matrix, vector, and set, respectively.
The notation used in the paper is summarized in Table \ref{Table:symbols}.

\subsection{Fading $2$-User $2$-Hop Networks} \label{subsec:2_2_2}
\begin{figure}[t!]
\begin{center}
\includegraphics[scale=1]{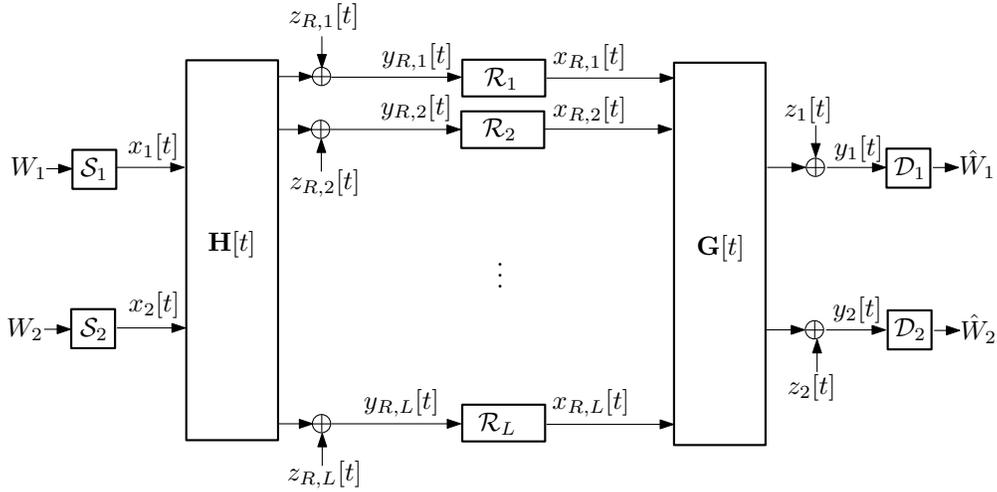}
\end{center}
\vspace{-0.15in}
\caption{Gaussian $2$-user $2$-hop network with $L$ relays.}
\label{fig:system_model}
\vspace{-0.1in}
\end{figure}

We study the $2$-user $2$-hop network depicted in Fig. \ref{fig:system_model} in which each source wishes to transmit an independent message to its destination with the help of $L$ relays, where $L\ge 2$.
The input--output relation of the first hop at time $t$ is given by
\begin{equation}
\mathbf{y}_R[t]=\mathbf{H}[t]\mathbf{x}[t]+\mathbf{z}_R[t],
\label{eq:in_out_vec1}
\end{equation}
where
\begin{equation}
\mathbf{H}[t]=
\left[
\begin{array}{cc}
  h_{1,1}[t]   & h_{1,2}[t]  \\
  h_{2,1}[t]   & h_{2,2}[t]  \\
    \vdots & \vdots  \\
   h_{L,1}[t]  & h_{L,2}[t]
\end{array}
\right]
\end{equation}
is the $L\times 2$ dimensional  complex channel matrix of the first hop at time $t$, $\mathbf{y}_R[t]=[y_{R,1}[t],\cdots,y_{R,L}[t]]^T$ is the $L\times 1$ dimensional  received signal vector of the relays at time $t$, $\mathbf{x}[t]=[x_1[t],x_2[t]]^T$ is the $2\times 1$ dimensional transmit signal vector of the sources at time $t$, and $\mathbf{z}_R[t]=[z_{R,1}[t],\cdots,z_{R,L}[t]]^T$ is the $L\times 1$ dimensional  noise vector of the relays at time $t$.
Similarly, the input--output relation of the second hop at time $t$ is given by
\begin{equation}
\mathbf{y}[t]=\mathbf{G}[t]\mathbf{x}_R[t]+\mathbf{z}[t],
\label{eq:in_out_vec2}
\end{equation}
where
\begin{equation}
\mathbf{G}[t]=
\left[
\begin{array}{cccc}
  g_{1,1}[t] &g_{1,2}[t]& \cdots & g_{1,L}[t]  \\
   g_{2,1}[t] &g_{2,2}[t]& \cdots &g_{2,L}[t]
\end{array}
\right]
\end{equation}
is the the $2\times L$ dimensional complex channel matrix of the second hop at time $t$,
$\mathbf{y}[t]=[y_1[t],y_2[t]]^T$ is the $2\times 1$ dimensional received signal vector of the destinations at time $t$, $\mathbf{x}_R[t]=[x_{R,1}[t],\cdots,x_{R,L}[t]]^T$ is the $L\times 1$ dimensional transmit signal vector of the relays at time $t$, and $\mathbf{z}[t]=[z_1[t],z_2[t]]^T$ is the $2\times 1$ dimensional noise vector of the destinations at time $t$.
We assume that the elements of $\mathbf{z}_R[t]$ and $\mathbf{z}[t]$ are  i.i.d. drawn from $\mathcal{CN}(0,1)$.
Each source and relay should satisfy the average power constraint $P$, i.e., $\E[|x_i[t]|^2]\leq P$ for $i\in\{1,2\}$ and $\E[|x_{R,j}[t]|^2]\leq P$ for $j\in\{1,\cdots,L\}$.

We assume that \emph{channel coefficients are i.i.d. drawn from a continuous distribution $f(x)$, $x\in\mathbb{C}$, and vary independently over time}.
Without loss of generality, we assume that $\E[|h_{i,j}[t]|^2]=1$ and $\E[|g_{j,i}[t]|^2]=1$ for all $i\in\{1,\cdots,L\}$ and $j\in\{1,2\}$.
We further assume that the sources do not know any channel state information (CSI) and the relays and the destinations know global CSI.
That is, at time $t$, each relay and destination knows $\mathbf{H}[t]$ and $\mathbf{G}[t]$.

\subsection{Ergodic Sum Capacity}
Based on the network model, we consider a set of length-$n$ block codes.
Let $W_i$ be the message of source $i$ uniformly distributed over $\{1,\cdots,2^{nR_i}\}$, where $R_i$ is the rate of source $i$.
A rate pair $(R_1,R_2)$ is said to be \emph{achievable} if there exists a sequence of $(2^{nR_1},2^{nR_2},n)$ codes such that the probabilities of error for $W_1$ and $W_2$ converge to zero as $n$ increases.
Notice that since channel coefficients are i.i.d. varying over time, an achievable rate pair $(R_1,R_2)$ is given as in the ergodic sense, i.e., the expectation over random channel coefficients.
The ergodic sum capacity $C_{\operatorname{sum}}$ is defined as the maximum achievable ergodic sum rate.
Unless otherwise specified, an achievable sum rate or the sum capacity in this paper mean an achievable ergodic sum rate or the ergodic sum capacity, respectively.

\section{Main Results} \label{sec:main_results}
In this section, we first introduce our main results.
Let $M:=\lfloor \frac{L}{2}\rfloor$.
As will be explained in Section \ref{sec:ergodic_in}, we only use $2M$ relays among the total number $L$ of relays for the achievability.
That is, the achievability is based on an even number of relays.
Without loss of generality, we assume that relay $1$ to relay $2M$ are used for relaying.
The achievability follows from ergodic interference neutralization based on amplify-and-forward relaying in which $2M$ relays are partitioned into $M$ pairs and interference is neutralized separately by each pair of relays.
In order to describe the proposed ergodic interference neutralization and its achievable sum rate, for $m\in\{1,\cdots,M\}$, we denote
\begin{equation}
\mathbf{H}_m[t]:=
\left[
\begin{array}{cc}
  h_{2m-1,1}[t] &  h_{2m-1,2}[t]  \\
   h_{2m,1}[t] & h_{2m,2}[t]
\end{array}
\right]
\end{equation}
and
\begin{equation}
\mathbf{G}_m[t]:=
\left[
\begin{array}{cc}
  g_{1,2m-1}[t] &  g_{2,2m-1}[t]  \\
   g_{1,2m}[t] & g_{2,2m}[t]
\end{array}
\right],
\end{equation}
which are the $2\times 2$ dimensional channel matrices at time $t$ from the sources to relays $2m-1$ and $2m$ and from relays $2m-1$ and $2m$ to the destinations, respectively.

\subsection{Achievable Sum Rate}
The following theorem states an achievable symmetric rate of the fading $2$-user $2$-hop network.

\vspace{0.1in}
\begin{theorem} \label{th:sum_rate}
For the fading $2$-user $2$-hop network with $L$ relays,
\begin{equation} \label{eq:achievable_rate}
R_i=\E\left[\log \left(1+\frac{P\gamma^2\left(\sum_{m=1}^M|\det(\mathbf{H}_m)|\right)^2}{1+\sigma^2_{\operatorname{AF},i}}\right)\right]
\end{equation}
is achievable for $i\in\{1,2\}$, where $M=\lfloor \frac{L}{2}\rfloor$, $\gamma=\sqrt{\frac{P}{1+2P}}$, $\sigma^2_{\operatorname{AF},i}=\gamma^2 \sum_{m=1}^M(|h_{2(m-1)+3-i,3-i}|^2+|h_{2(m-1)+i,3-i}|^2)$, and the expectation is over the channel coefficients.
\end{theorem}
\vspace{0.1in}
\begin{proof}
The proof is in Section \ref{sec:ergodic_in}.
\end{proof}
\vspace{0.1in}

The most important aspect is that there is no residual interference after ergodic interference neutralization, meaning that interference can completely be neutralized at finite SNR.
Moreover, from the block-wise coherent combining gain shown as $(\sum_{m=1}^M |\det(\mathbf{H}_m)|)^2$ in \eqref{eq:achievable_rate}, the received signal power increases as the number of pairs $M$ increases.
Although there is noise amplification due to amplify-and-forward relaying given as $\sigma^2_{\operatorname{AF},i}$ in \eqref{eq:achievable_rate}, this additional noise results in a constant number of bits/sec/Hz loss for a broad class of channel distributions, which will be proved in Section \ref{sec:appro_cap}.

For notational convenience, let
\begin{equation} \label{eq:R_in}
R_{\operatorname{in}}:=\sum_{i=1}^2\E\left[\log \left(1+\frac{P\gamma^2\left(\sum_{m=1}^M|\det(\mathbf{H}_m)|\right)^2}{1+\sigma^2_{\operatorname{AF},i}}\right)\right],
\end{equation}
which is the achievable sum rate from Theorem \ref{th:sum_rate}.
For comparison, we consider the ergodic capacity of the multiple-input multiple-output (MIMO) channel from the sources to the relays, that is
\begin{equation} \label{eq:R_mimo}
R_{\operatorname{mimo}}:=\E\left[\log\det(\mathbf{I}+P\mathbf{H}\mathbf{H}^\dagger)\right].
\end{equation}
Since the channel coefficients are i.i.d. and the sources do not know CSI, $C_{\operatorname{sum}}$ is upper bounded by $R_{\operatorname{mimo}}$ \cite{Telatar:99}.
The following example illustrates $R_{\operatorname{in}}$ and $R_{\operatorname{mimo}}$ for i.i.d. Rayleigh fading, i.e., $f(x)$ follows $\mathcal{CN}(0,1)$.

\begin{figure}[t!]
\begin{center}
\includegraphics[scale=0.65]{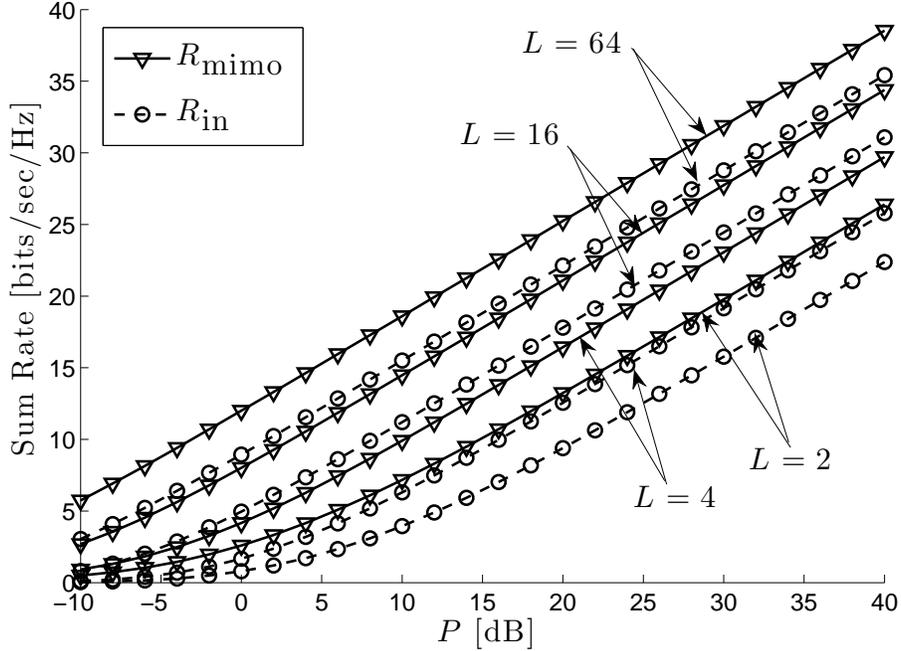}
\end{center}
\vspace{-0.15in}
\caption{The achievable sum rate $R_{\operatorname{in}}$ and its upper bound $R_{\operatorname{mimo}}$ for i.i.d. Rayleigh fading when $L=2,4,16,64$.}
\label{fig:sum_rate_rayleigh}
\vspace{-0.1in}
\end{figure}

\vspace{0.1in}
\begin{example}[Sum rate: Rayleigh fading] \label{ex:sum_rate_rayleigh}
Figure \ref{fig:sum_rate_rayleigh} plots $R_{\operatorname{in}}$ and $R_{\operatorname{mimo}}$ for i.i.d. Rayleigh fading.
Two important aspects can be observed in the figure.
First, for a fixed number of relays $L$, the sum rate gap $R_{\operatorname{mimo}}-R_{\operatorname{in}}$ appears to be upper bounded by some constant independent of power $P$, which suggests that the proposed ergodic interference neutralization can achieve the ergodic sum capacity to within a constant number of bits/sec/Hz independent of $P$.
Second, for a fixed $P$, the sum rate gap $R_{\operatorname{mimo}}-R_{\operatorname{in}}$ appears to decrease with increasing $L$, which suggests that this approximate capacity characterization can be tightened as the number of relays $L$ increases.
\end{example}
\vspace{0.1in}

Both observations in Example \ref{ex:sum_rate_rayleigh} are established in this paper and shown to hold beyond the case of Rayleigh fading for any fading model for which $f(x)$ is only a function of $|x|$.
The following two subsections describe our approximate capacity results characterizing the ergodic sum capacity to within a constant number of bits/sec/Hz, independent of $P$.

\subsection{Approximate Ergodic Sum Capacity for $L=2$} \label{subsec:approximate_cap_1}
In this subsection, we assume $L=2$.
We first consider i.i.d. uniform phase fading in which $h_{i,j}[t]=\exp(\jmath\theta_{i,j}[t])$ and $g_{j,i}[t]=\exp(\jmath\varphi_{j,i}[t])$, where $\theta_{i,j}[t]$ and $\varphi_{j,i}[t]$ are uniformly distributed over $[0,2\pi)$ for all $i,j\in\{1,2\}$.
Although uniform phase fading violates the channel assumption in Section \ref{subsec:2_2_2}, i.e., $f(x)$ is continuous over $x\in\mathbb{C}$, we can slightly modify the proposed ergodic interference neutralization and show that Theorem \ref{th:sum_rate} still holds.
The detailed modification is given in Appendix I.
The following theorem characterizes an approximate ergodic sum capacity for i.i.d. uniform phase fading.

\vspace{0.1in}
\begin{theorem}\label{th:constant_gap_unit_amp}
Consider the fading $2$-user $2$-hop network with $L=2$ relays.
If $h_{i,j}[t]=\exp(\jmath\theta_{i,j}[t])$ and $g_{j,i}[t]=\exp(\jmath\varphi_{j,i}[t])$, where $\theta_{i,j}[t]$ and $\varphi_{j,i}[t]$ are uniformly distributed over $[0,2\pi)$ for all $i,j\in\{1,2\}$, then
\begin{equation}
C_{\operatorname{sum}}-R_{\operatorname{in}}\leq4
\end{equation}
for any $P>0$.
\end{theorem}
\vspace{0.1in}
\begin{proof}
The proof is in Section \ref{subsec:l_2}.
\end{proof}
\vspace{0.1in}


\begin{figure}[t!]
\begin{center}
\includegraphics[scale=0.65]{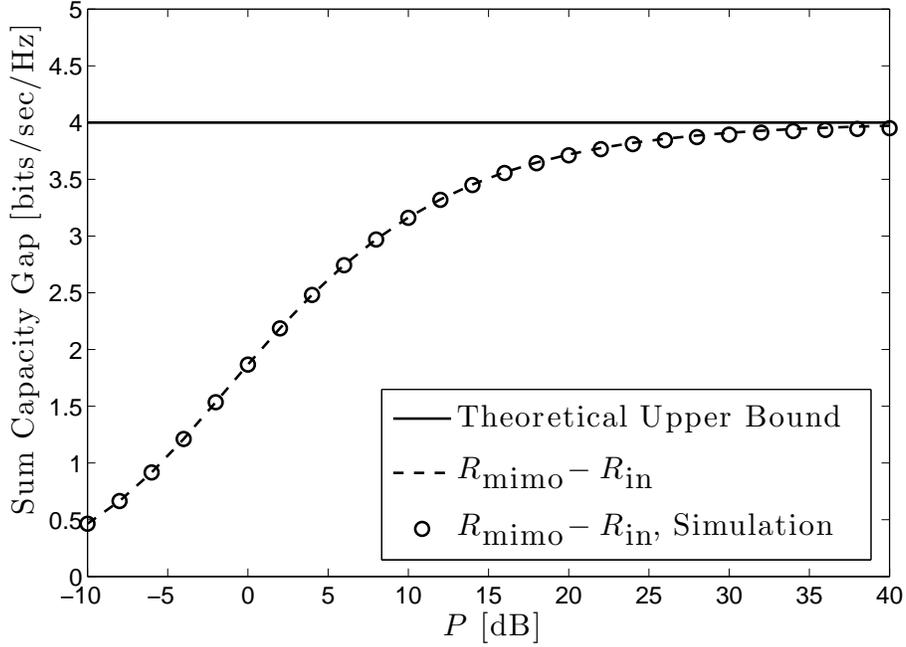}
\end{center}
\vspace{-0.15in}
\caption{Gap from the sum capacity for i.i.d. uniform phase fading when $L=2$.}
\label{fig:gap}
\vspace{-0.1in}
\end{figure}

\vspace{0.1in}
\begin{example}[Gap for $L=2$: Uniform phase fading]
Figure \ref{fig:gap} plots $R_{\operatorname{mimo}}-R_{\operatorname{in}}$ with respect to $P$ for i.i.d. uniform phase fading (the closed forms  of $R_{\operatorname{mimo}}$ and $R_{\operatorname{in}}$ are given by \eqref{eq:lower_unit_amp1} and \eqref{eq:upper_unit_amp1}, respectively).
As proved by Theorem \ref{th:constant_gap_unit_amp}, the proposed ergodic interference neutralization achieves $C_{\operatorname{sum}}$ to within 4 bits/sec/Hz for i.i.d. uniform phase fading.
This theoretical gap coincides with the actual gap $R_{\operatorname{mimo}}-R_{\operatorname{in}}$ at high SNR, i.e., $\lim_{P\to\infty}\{R_{\operatorname{mimo}}-R_{\operatorname{in}}\}=4$.
\end{example}
\vspace{0.1in}

Based on the bounding techniques used in proving Theorem \ref{th:constant_gap_unit_amp}, we characterize an approximate ergodic sum capacity for a  class of channel distributions satisfying that $f(x)$ is only a function of $|x|$.
Specifically, for a given set of channel amplitudes, we first upper bound the gap $R_{\operatorname{mimo}}-R_{\operatorname{in}}$ by averaging out the effect of phase fading.
Then we further apply additional bounding techniques  to obtain an upper bound, independent of power $P$.

\vspace{0.1in}
\begin{theorem} \label{th:constant_gap_arb_amp_iid}
Consider the fading $2$-user $2$-hop network with $L=2$ relays.
If $f(x)$ is only a function of $|x|$, then
\begin{align}
C_{\operatorname{sum}}-R_{\operatorname{in}}\le2\E\left[\log\left(\frac{\sqrt{A}(A+B^2)}{B(A+\sqrt{A^2-G^2})}\right)\right]+2
\end{align}
for any $P> 0$, where
\begin{align} \label{eq:def_abg}
A&=|h_{1,1}|^2|h_{2,2}|^2+|h_{1,2}|^2|h_{2,1}|^2,\nonumber\\
B&=|h_{1,1}|^2+|h_{2,1}|^2+2,\nonumber\\
G&=2|h_{1,1}||h_{1,2}||h_{2,1}||h_{2,2}|,
\end{align}
and the expectation is over the channel coefficients.
\end{theorem}
\vspace{0.1in}
\begin{proof}
The proof is in Section \ref{subsec:l_2}.
\end{proof}
\vspace{0.1in}

The presented gap in Theorem \ref{th:constant_gap_arb_amp_iid} only depends on the amplitude distribution of channel coefficients, which provides universal performance guarantee regardless of power $P$.
The following example evaluates the presented gap for i.i.d. Rayleigh fading.

\begin{figure}[t!]
\begin{center}
\includegraphics[scale=0.65]{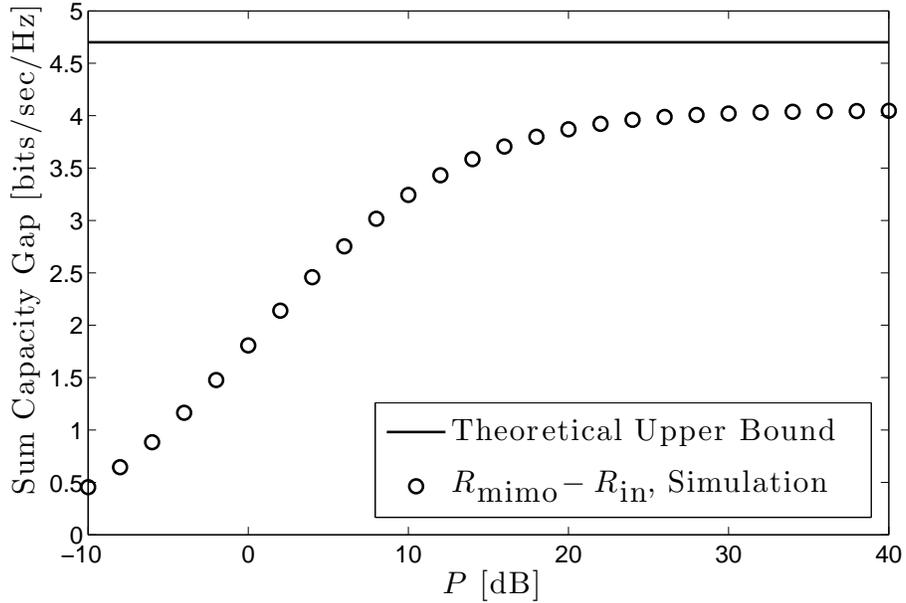}
\end{center}
\vspace{-0.15in}
\caption{Gap from the sum capacity for i.i.d. Rayleigh fading when $L=2$.}
\label{fig:gap_uniform_arb_amp_iid}
\vspace{-0.1in}
\end{figure}

\vspace{0.1in}
\begin{example}[Gap for $L=2$: Rayleigh fading] \label{ex:rayleigh_2}
Figure \ref{fig:gap_uniform_arb_amp_iid} plots $R_{\operatorname{mimo}}-R_{\operatorname{in}}$ with respect to $P$ and also plots its upper bound in Theorem \ref{th:constant_gap_arb_amp_iid} for i.i.d. Rayleigh fading.
Since there is no closed form, we evaluate the bound in Theorem \ref{th:constant_gap_arb_amp_iid} by simulation, which approximately provides $4.7$ bits/sec/Hz gap.
Simulation result shows that the proposed scheme achieves at least $71\%$, $79\%$, $84\%$, $87\%$, and $89\%$ percent of the ergodic sum capacity at SNR $20$, $30$, $40$, $50$, and $60$ dB, respectively.  
\end{example}

\subsection{Approximate Ergodic Sum Capacity as $L\to \infty$} \label{subsec:approximate_cap_infty}
In this subsection, we focus on an approximate ergodic sum capacity as the number $L$ of relays increases.
Again, we first consider i.i.d. uniform phase fading and then consider a class of channel distributions satisfying that $f(x)$ is only a function of $|x|$.

\vspace{0.1in}
\begin{theorem} \label{th:constant_gap_unit_amp_L}
Consider the fading $2$-user $2$-hop network with $L$ relays.
If $h_{i,j}[t]=\exp(\jmath\theta_{i,j}[t])$, $g_{j,i}[t]=\exp(\jmath\varphi_{j,i}[t])$, and $\theta_{i,j}[t]$ and $\varphi_{j,i}[t]$ are uniformly distributed over $[0,2\pi)$ for all $i\in\{1,\cdots,L\}$ and $j\in\{1,2\}$, then
\begin{align}
\lim_{L\to\infty}\{C_{\operatorname{sum}}-R_{\operatorname{in}}\}&\leq 4\log \pi-4
\end{align}
for any $P>0$.
\end{theorem}
\vspace{0.1in}
\begin{proof}
The proof is in Section \ref{subsec:l_infty}.
\end{proof}
\vspace{0.1in}

\begin{figure}[t!]
\begin{center}
\includegraphics[scale=0.65]{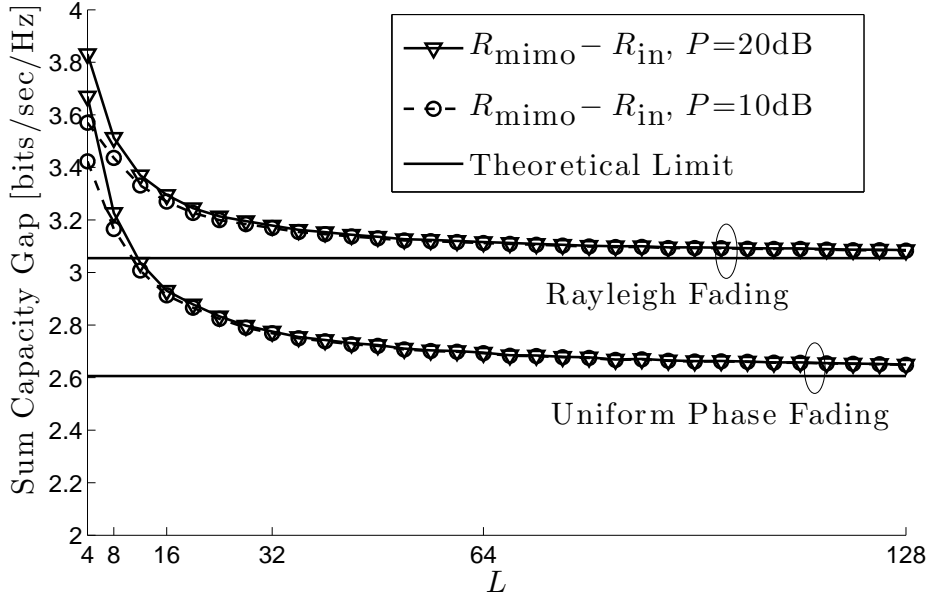}
\end{center}
\vspace{-0.15in}
\caption{Gap from the sum capacity with respect to the number of relays.}
\label{fig:gap_L_uniform}
\vspace{-0.1in}
\end{figure}

\begin{example}[Gap as $L\to\infty$: Uniform phase fading]{}
Figure \ref{fig:gap_L_uniform} plots the gap $R_{\operatorname{mimo}}-R_{\operatorname{in}}$ for i.i.d. uniform phase fading with respect to $L$.
As shown in the figure, this gap decreases as $L$ increases and eventually converges to $4\log \pi-4$ (approximately $2.6$) regardless of $P$, which was proved in Theorem \ref{th:constant_gap_unit_amp_L}. Therefore the proposed ergodic interference neutralization characterizes $C_{\operatorname{sum}}$ to within $4\log \pi-4$ bits/sec/Hz in the limit of large $L$.
Compared to $4$ bits/sec/Hz, the sum capacity gap for $L=2$ in Theorem \ref{th:constant_gap_unit_amp}, the result shows that the sum capacity gap can be tightened as $L$ increases.
\end{example}

\vspace{0.1in}
\begin{theorem}\label{th:constant_gap_arb_L}
Consider the fading $2$-user $2$-hop network with $L$ relays.
If $f(x)$ is only a function of $|x|$, then
\begin{align}
\lim_{L\to\infty}\{C_{\operatorname{sum}}-R_{\operatorname{in}}\}\leq 4-4\log\left(\E[|\det(\mathbf{H}_1)|]\right)
\end{align}
for any $P>0$.
\end{theorem}
\vspace{0.1in}
\begin{proof}
The proof is in Section \ref{subsec:l_infty}.
\end{proof}

\vspace{0.1in}
\begin{example}[Gap as $L\to\infty$: Rayleigh fading]{}
Figure \ref{fig:gap_L_uniform} plots $R_{\operatorname{mimo}}-R_{\operatorname{in}}$ for i.i.d. Rayleigh fading with respect to $L$.
That is, $f(x)$ follows $\mathcal{CN}(0,1)$. For this case, it can be shown that $\E[|\det(\mathbf{H}_1)|]=\frac{3\pi}{8}$ and, thus, the theoretical limit in Theorem \ref{th:constant_gap_arb_L} leads $4-4\log( \frac{3\pi}{8})$ (approximately $3.1$).
The detailed proof of $\E[|\det(\mathbf{H}_1)|]=\frac{3\pi}{8}$ is in Appendix II.
As shown in the figure, $R_{\operatorname{mimo}}-R_{\operatorname{in}}$ quickly converges to the theoretical limit as $L$ increases.
Considering that  the sum capacity gap is approximately given by $4.7$ bits/sec/Hz when $L=2$ (Theorem \ref{th:constant_gap_arb_amp_iid} and Example \ref{ex:rayleigh_2}), the sum capacity gap can be tightened as $L$ increases.
\end{example}

\subsection{Approximate Ergodic Capacity for Fading Interference Channel} \label{subsec:approximate_cap_ic}
We notice that a similar analysis used in Theorems \ref{th:constant_gap_unit_amp} and \ref{th:constant_gap_arb_amp_iid} is applicable to show an approximate ergodic capacity for fading $K$-user interference channel.
The achievability follows from ergodic interference alignment in \cite{Nazer11:09}. 
Assuming that all sources employ uniform power allocation across time, we show that ergodic interference alignment characterizes an approximate ergodic per-user capacity, i.e., ergodic sum capacity divided by $K$, for a broad class of channel distributions.
The detailed statement is given in Theorem \ref{th:eia} in Appendix III.
For i.i.d. Rayleigh fading, for instance, our analysis characterizes the ergodic per-user capacity to within $\frac{1}{2}\log 6$ bits/sec/Hz (approximately $1.3$ bits/sec/Hz).

\section{Ergodic Interference Neutralization} \label{sec:ergodic_in}

For the achievability, we propose ergodic interference neutralization using an even number of relays.
Let $M:=\lfloor \frac{L}{2}\rfloor$.
Then we can choose $2M$ relays among the total number $L$ of relays and apply the proposed ergodic interference neutralization by using these $2M$ relays.
For simplicity, we assume $L$ is even in the rest of this section.
That is, $L=2M$.

\subsection{High-Level View}
Before the detailed description and analysis, we begin by providing a high-level view of the proposed ergodic interference neutralization.
Consider length-$n$ sequences of matrices $\{\mathbf{H}[t]\}_{t=1}^n$ and $\{\mathbf{G}[t]\}_{t=1}^{n}$, drawn i.i.d. according to a certain
probability density function. We partition these sequences
judiciously into pairs of matrices $(\mathbf{H}[t_1], \mathbf{G}[t_2])$ such that
$\mathbf{G}[t_2]$ and $F(\mathbf{H}[t_1])$ are \emph{almost equal}, where $F(\cdot)$ is a cleverly chosen
mapping to be discussed below. The main
argument is that by considering a longer and longer sequence of
matrices, we can make these two matrices arbitrarily close. The formal and
technical details of this argument can be found in Sections \ref{subsec:pairing} and \ref{subsec:achievable_rate}.
For notational convenience, we introduce the notation $\mathbf{G}[t_2]\simeq F(\mathbf{H}[t_1])$ for the two matrices that are almost equal.

\begin{figure}[t!]
\begin{center}
\includegraphics[scale=1]{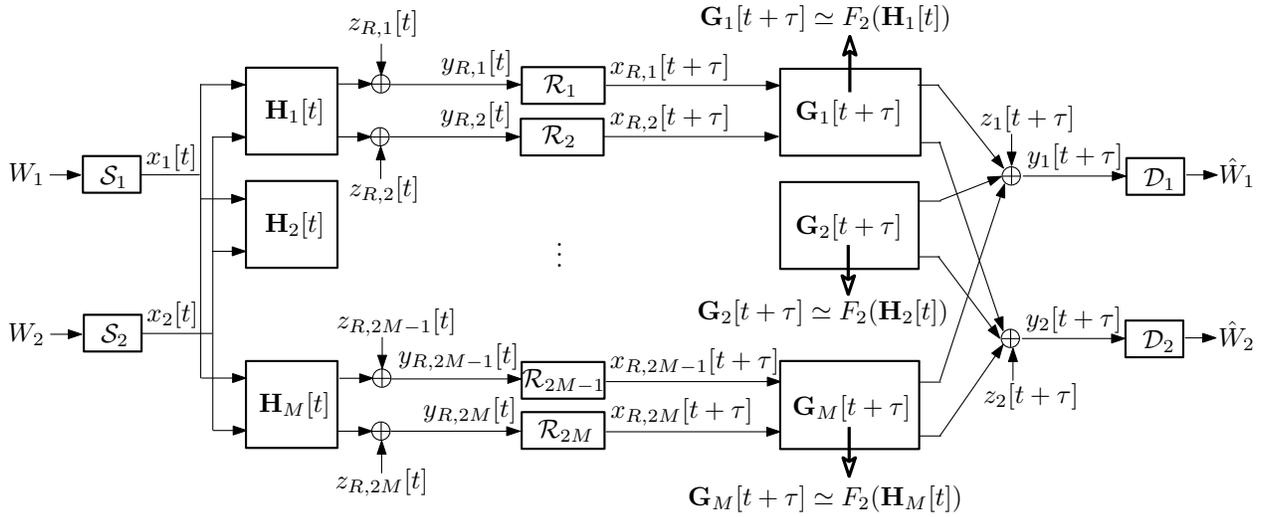}
\end{center}
\vspace{-0.15in}
\caption{Block-wise ergodic interference neutralization based on amplify-and-forward relaying.}
\label{fig:block_in}
\vspace{-0.1in}
\end{figure}

As pointed out in \cite{Jeon2:11}, a simple amplify-and-forward scheme with an appropriate delay $\tau\in\mathbb{Z}_+$ can neutralize interference by letting $\mathbf{G}[t+\tau]\mathbf{H}[t]$ approximately a diagonal matrix with non-zero diagonal elements.
To satisfy this condition, we first partition $L$ relays into $M=\frac{L}{2}$ pairs and neutralize interference separately by each pair of relays.
Figure \ref{fig:block_in} illustrates the main idea of the proposed scheme.
For $\mathbf{A}=\{a_{i,j}\}\in\mathbb{C}^{2\times 2}$, define
\begin{equation} \label{eq:F_2}
F_2(\mathbf{A}):=
\left[
\begin{array}{cc}
  a_{2,2} & a_{1,2}  \\
   a_{2,1} &a_{1,1}
\end{array}
\right].
\end{equation}
The relays then amplify and forward with delay $\tau$ such that $\mathbf{G}_m[t+\tau]\simeq F_2(\mathbf{H}_m[t])$ for all $m\in\{1,\cdots,M\}$.
For relaying, relays $2m-1$ and $2m$ amplify and forward with the amplification factors $\gamma\frac{\det(\mathbf{H}_m[t])^*}{|\det(\mathbf{H}_m[t]|}$ and $-\gamma\frac{\det(\mathbf{H}_m[t])^*}{|\det(\mathbf{H}_m[t]|}$, respectively.
Here $\gamma=\sqrt{\frac{P}{1+2P}}$ is needed to satisfy the average power constraint $P$.
Then the effective channel matrix of the $m$th pair is given by
\begin{align}
\gamma\frac{\det(\mathbf{H}_m[t])^*}{|\det(\mathbf{H}_m[t])|}\mathbf{G}_m[t+\tau]\left[
\begin{array}{cc}
  1 & 0  \\
   0 &-1
\end{array}
\right]
\mathbf{H}_m[t]&\simeq
\gamma\frac{\det(\mathbf{H}_m[t])^*}{|\det(\mathbf{H}_m[t])|}F_2(\mathbf{H}_m[t])\left[
\begin{array}{cc}
  1 & 0  \\
   0 &-1
\end{array}
\right]
\mathbf{H}_m[t]\nonumber\\
&=\gamma|\det(\mathbf{H}_m[t])|\left[\begin{array}{cc}
  1 & 0  \\
   0 &-1
\end{array}\right].
\end{align}
As a consequence, the effective channel gain from each source to its destination is approximately given by $\gamma^2(\sum_{m=1}^M|\det(\mathbf{H}_m[t])|)^2$, as can be seen in the rate expression in Theorem \ref{th:sum_rate}.
One can easily show that the additional noise power at destination $i$ due to this amplify-and-forward relaying is given as $\sigma^2_{\operatorname{AF},i}$, as shown in the rate expression in Theorem \ref{th:sum_rate}.
Lastly, since the probability density functions of the paired channel states are the same, i.e.,
\begin{equation}
f_{\mathbf{H}[t]}([\mathbf{H}_1^T,\cdots,\mathbf{H}_M^T]^T)=f_{\mathbf{G}[t]}([F_2(\mathbf{H}_1),\cdots,F_2(\mathbf{H}_M)]),
\end{equation}
almost all channel instances can be utilized for this ergodic pairing as the block length $n$ increases.
Hence, the ergodic rate in Theorem \ref{th:sum_rate} is achievable in the limit of large $n$.

There are two crucial facts to be observed: 1) the intended signal power received at each destination is non-zero while the interference power decreases arbitrarily close to zero at any finite power $P$; 2) the intended signal power received at each destination increases quadratically with increasing $L$.
These facts make approximate capacity characterization possible for a broad class of channel distributions.

Although finding a pair of channel instances having exact prescribed values is impossible, such a pairing can be done approximately by partitioning the channel space of each hop and then pairing the partitioned channel spaces between the first and second hops.
In the following subsection, we first explain channel space partition and pairing and then explain the detailed scheme.

\subsection{Block-Wise Ergodic Interference Neutralization}  \label{subsec:pairing}

\subsubsection{Partitioning and pairing of channel space}
\begin{figure}[t!]
\begin{center}
\includegraphics[scale=0.9]{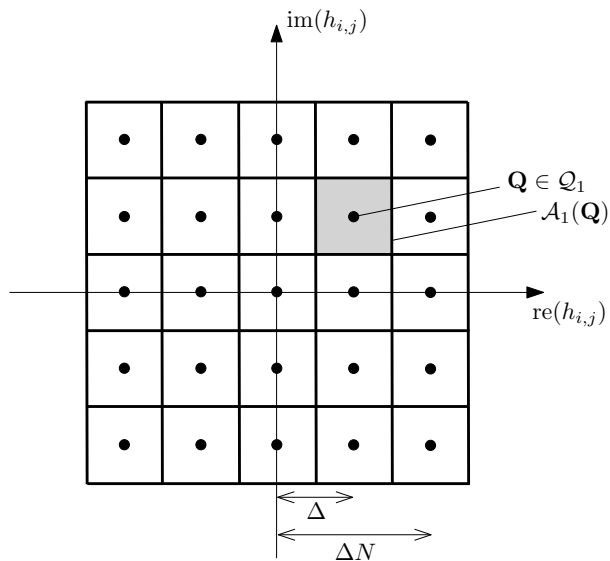}
\end{center}
\vspace{-0.15in}
\caption{Channel space partitioning with respect to the channel coefficient $h_{i,j}\in \mathbb{C}$.}
\label{fig:quan_1}
\vspace{-0.1in}
\end{figure}

We partition the channel space of each hop, i.e, $\mathbb{C}^{2M\times 2}$ space for the first hop and $\mathbb{C}^{2\times 2M}$ space for the second hop.
First, consider the channel space of the first hop $\mathbb{C}^{2M\times 2}$.
For $N\in\mathbb{Z}_+$ and $\Delta>0$, define
\begin{align} \label{eq:Q_1}
\mathcal{Q}_1:=&\big\{\mathbf{A}\in\Delta(\mathbb{Z}^{2M\times 2}+\jmath\mathbb{Z}^{2M\times 2})\big||\mbox{re}(a_{i,j})|\leq \Delta N,|\mbox{im}(a_{i,j})|\leq \Delta N\nonumber\\
&\mbox{ for all } i\in\{1,\cdots,2M\}\mbox{ and }j\in\{1,2\}\big\},
\end{align}
where $\mathbf{A}=\{a_{i,j}\}$.
Here, $N$ and $\Delta$ are related to the number of quantization points and the quantization interval.
For a quantized channel matrix $\mathbf{Q}\in\mathcal{Q}_1$, define
\begin{align} \label{eq:A_1_Q}
\mathcal{A}_1(\mathbf{Q}):=&\bigg\{\mathbf{A}\in{\mathbb{C}^{2M\times 2}}\big|-\frac{\Delta}{2}\le \mbox{re}(a_{i,j})-\mbox{re}(q_{i,j})<\frac{\Delta}{2}\mbox{ and }-\frac{\Delta}{2}\le \mbox{im}(a_{i,j})-\mbox{im}(q_{i,j})<\frac{\Delta}{2}\nonumber\\
&\mbox{ for all }i\in\{1,\cdots,2M\}\mbox{ and }j\in\{1,2\}\bigg\},
\end{align}
where $\mathbf{A}=\{a_{i,j}\}$ and $\mathbf{Q}=\{q_{i,j}\}$.
Figure \ref{fig:quan_1} illustrates the channel space partitioning with respect to $h_{i,j}\in \mathbb{C}$.
We can define $\mathcal{Q}_2$ and $\mathcal{A}_2(\mathbf{Q})$ for the second hop as the same manner in \eqref{eq:Q_1} and \eqref{eq:A_1_Q} by substituting $\mathbf{A}\in\Delta(\mathbb{Z}^{2\times 2M}+\jmath\mathbb{Z}^{2\times 2M})$ and $\mathbf{A}\in{\mathbb{C}^{2\times 2M}}$, respectively.
We will only use the first-hop channel instances in $\cup_{\mathbf{Q}\in\mathcal{Q}_1}\mathcal{A}_1(\mathbf{Q})$ and the second-hop channel instances in $\cup_{\mathbf{Q}\in\mathcal{Q}_2}\mathcal{A}_2(\mathbf{Q})$ for transmission.

Now consider the channel space pairing between  $\mathcal{A}_1(\mathbf{Q})$ and $\mathcal{A}_2(\mathbf{Q})$.
For $\mathbf{A}\in\mathbb{C}^{2M\times 2}$, define
\begin{equation}
F(\mathbf{A}):=[F_2(\mathbf{A}_1),F_2(\mathbf{A}_2),\cdots,F_2(\mathbf{A}_M)],
\end{equation}
where $\mathbf{A}=[\mathbf{A}_1^T,\mathbf{A}_2^T,\cdots,\mathbf{A}_M^T]^T$ and the definition of $F_2(\cdot)$ is given by \eqref{eq:F_2}.
For  $\mathbf{H}[t]\in \mathcal{A}_1(\mathbf{Q})$, the relays will amplify and forward with delay $\tau\in\mathbb{Z}_+$ satisfying $\mathbf{G}[t+\tau]\in\mathcal{A}_2(F(\mathbf{Q}))$. Hence the channel subspace $ \mathcal{A}_1(\mathbf{Q})$ of the first hop is paired with the channel subspace $ \mathcal{A}_2(F(\mathbf{Q}))$ of the second hop.
The detailed transmission scheme is given in the following subsection.

\subsubsection{Transmission scheme}
We first divide a length-$n$ block into $B$ sub-blocks having length $n_B=\frac{n}{B}$ each.
At the first sub-block, the sources transmit their first messages to the relays (the relays do not transmit).
At the $b$th sub-block, $b\in\{2,\cdots,B-1\}$, the sources transmit their $b$th messages to the relays and the relays amplify and forward the received signals of the $(b-1)$th sub-block to the destinations. At the last sub-block, the relays amplify and forward the received signals of the $(B-1)$th sub-block to the destinations (the sources do not transmit).
Hence, the number of effective sub-blocks is equal to $B-1$.
Since we can set both $n_B$ and $B$ as large as desired as $n$ increases, the fractional rate loss $\frac{1}{B}$ becomes negligible as $n$ increases.
For simplicity, we describe the proposed scheme based on the first message transmission and omit the sub-block index.

For $\mathbf{Q}\in\mathcal{Q}_1$, define $\mathcal{T}_1(\mathbf{Q}):=\big\{t\in\{1,\cdots,n_B\}\big|\mathbf{H}[t]\in\mathcal{A}_1(\mathbf{Q})\big\}$, which is the set of time indices of the first hop whose channel instances belong to $\mathcal{A}_1(\mathbf{Q})$.
Similarly, for $\mathbf{Q}\in\mathcal{Q}_2$, $\mathcal{T}_2(\mathbf{Q}):=\left\{t\in\{n_B+1,\cdots,2n_B\}\big|\mathbf{G}[t]\in\mathcal{A}_2(\mathbf{Q})\right\}$, which is the set of time indices of the second hop whose channel instances belong to $\mathcal{A}_2(\mathbf{Q})$.
The encoding, relaying, and decoding are as follows.
\begin{itemize}
\item (Encoding)
The sources transmit their messages using Gaussian codebook with length $n_B$ and average power $P$.
\item (Relaying)
For all $\mathbf{Q}\in\mathcal{Q}_1$, the relays amplify and forward their received signals that were received during $\mathcal{T}_1(\mathbf{Q})$ using the time indices in $\mathcal{T}_2(F(\mathbf{Q}))$.
Specifically, for $t_1\in\mathcal{T}_1(\mathbf{Q})$, the transmit signal vector of the relays is given by $\mathbf{x}_R[t_2]=\mathbf{\Gamma}\mathbf{y}_R[t_1]$, where $t_2\in\mathcal{T}_2(F(\mathbf{Q}))$.
Here
\begin{equation}
\mathbf{\Gamma} = \left[
                          \begin{array}{cccc}
                            \gamma\frac{\det(\mathbf{Q}_1)^*}{|\det(\mathbf{Q}_1)|}\mathbf{\Lambda} &\mathbf{0} &\cdots & \mathbf{0} \\
                           \mathbf{0} & \gamma\frac{\det(\mathbf{Q}_2)^*}{|\det(\mathbf{Q}_2)|}\mathbf{\Lambda} & &\vdots\\
                           \vdots &  &\ddots &\\
                           \mathbf{0} & \cdots & &\gamma\frac{\det(\mathbf{Q}_M)^*}{|\det(\mathbf{Q}_M)|}\mathbf{\Lambda}
                          \end{array}
                        \right],
\end{equation}
$\gamma=\sqrt{\frac{P}{1+2P}}$, and $\mathbf{\Lambda}=[[1,0]^T[0,-1]^T]^T$, where $\mathbf{Q}=[\mathbf{Q}_1^T,\cdots,\mathbf{Q}_M^T]^T$ and $\mathbf{0}$ denotes the $2\times2$ dimensional all-zero matrix.
\item(Decoding) The destinations decode their messages based on their received signals during $t\in\{n_B+1,\cdots,2n_B\}$.
\end{itemize}

\subsection{Achievable Rate Region} \label{subsec:achievable_rate}

In this subsection, we prove Theorem \ref{th:sum_rate}.
We first introduce the following two lemmas.

\vspace{0.1in}
\begin{lemma} \label{lemma:prob}
For any $\mathbf{Q}\in\mathcal{Q}_1$,
\begin{equation}
\P\left[\mathbf{H}[t]\in \mathcal{A}_1(\mathbf{Q})\right]=\P\left[\mathbf{G}[t]\in \mathcal{A}_2(F(\mathbf{Q}))\right].
\end{equation}
\end{lemma}
\vspace{0.1in}
\begin{proof}
Let $f_{\mathbf{H}[t]}(\cdot)$ and $f_{\mathbf{G}[t]}(\cdot)$ denote the probability density functions of $\mathbf{H}[t]$ and $\mathbf{G}[t]$, respectively. Then
\begin{align}
\P\left[\mathbf{H}[t]\in \mathcal{A}_1(\mathbf{Q})\right]&=\int_{\mathbf{A}\in\mathcal{A}_1(\mathbf{Q})}f_{\mathbf{H}[t]}(\mathbf{A})d\mathbf{A}\nonumber\\
&=\int_{\mathbf{A}\in\mathcal{A}_1(\mathbf{Q})}\prod_{i\in\{1,\cdots2M\},j\in\{1,2\}}f(a_{i,j})d\mathbf{A}\nonumber\\
&\overset{(a)}{=}\int_{\mathbf{A}\in\mathcal{A}_1(\mathbf{Q})}f_{\mathbf{G}[t]}(F(\mathbf{A}))d\mathbf{A}\nonumber\\
&\overset{(b)}{=}\int_{\mathbf{A}'\in\mathcal{A}_2(F(\mathbf{Q}))}f_{\mathbf{G}[t_2]}(\mathbf{A}')d\mathbf{A}'\nonumber\\
&=\P\left[\mathbf{G}[t]\in \mathcal{A}(F(\mathbf{Q}))\right],
\end{align}
where $\mathbf{A}=\{a_{i,j}\}$.
Here $(a)$ follows from the definition of $F(\mathbf{A})$ and $(b)$ follows by a change of variable $\mathbf{A}'=F(\mathbf{A})$ whose Jacobian is one and $\mathcal{A}_2(F(\mathbf{Q}))=\{F(\mathbf{A})|\mathbf{A}\in\mathcal{A}_1(\mathbf{Q})\}$.
Therefore Lemma \ref{lemma:prob} holds.
\end{proof}

\vspace{0.1in}
\begin{lemma}The probability that
\begin{equation}
\left|\frac{\mbox{card}(\mathcal{T}_1(\mathbf{Q}_1))}{n_B}-\P[\mathbf{H}[t]\in \mathcal{A}_1(\mathbf{Q}_1)]\right|\leq \delta
\end{equation}
and
\begin{equation}
\left|\frac{\mbox{card}(\mathcal{T}_2(\mathbf{Q}_2))}{n_B}-\P[\mathbf{G}[t]\in \mathcal{A}_2(\mathbf{Q}_2)]\right|\leq \delta
\end{equation}
for all $\mathbf{Q}_1\in \mathcal{Q}_1$ and $\mathbf{Q}_2\in \mathcal{Q}_2$  is greater than $1-(\mbox{card}(\mathcal{Q}_1)+\mbox{card}(\mathcal{Q}_2))/(2n_B \delta^2)$.
\label{lemma:large_num}
\end{lemma}
\vspace{0.1in}
\begin{proof}
We refer to Lemma 2.12 in \cite{Csiszar:81} for the proof.
\end{proof}
\vspace{0.1in}

Suppose that the sources transmit at time $t_1\in\mathcal{T}_1(\mathbf{Q})$ and the relays amplify and forward their received signals at time $t_2\in\mathcal{T}_2(F(\mathbf{Q}))$, where $\mathbf{Q}\in\mathcal{Q}_1$.
For this case, from (\ref{eq:in_out_vec1}) and (\ref{eq:in_out_vec2}), the received signal vector of the destinations is given by
\begin{equation} \label{eq:received_signal}
\mathbf{y}[t_2]=\mathbf{G}[t_2]\mathbf{\Gamma}\mathbf{H}[t_1]\mathbf{x}[t_1]+\mathbf{G}[t_2]\mathbf{\Gamma}\mathbf{z}_R[t_1]+\mathbf{z}[t_2],
\end{equation}
where we use $\mathbf{x}_R[t_2]=\mathbf{\Gamma}\mathbf{y}_R[t_1]$.
Denote $\mathbf{H}[t_1]=\mathbf{H}=[\mathbf{H}_1^T,\cdots\mathbf{H}_M^T]^T$ and $\mathbf{G}[t_2]=F(\mathbf{H})+\mathbf{\Delta}$, where $\mathbf{\Delta}=[\mathbf{\Delta}_1,\cdots,\mathbf{\Delta}_M]$ is the quantization error matrix with respect to $F(\mathbf{H})$.
From \eqref{eq:received_signal},
\begin{align}
\mathbf{y}[t_2]&=\left(\left(\gamma\sum_{m=1}^M|\det(\mathbf{H}_m)|\right)\mathbf{\Lambda}+\mathbf{\Delta}\mathbf{\Gamma}\mathbf{H}\right)\mathbf{x}[t_1]+(F(\mathbf{H})+\mathbf{\Delta})\mathbf{\Gamma}\mathbf{z}_R[t_1]+\mathbf{z}[t_2],
\end{align}
where we use $F(\mathbf{H})\mathbf{\Gamma}\mathbf{H}=(\gamma\sum_{m=1}^M|\det(\mathbf{H}_m)|)\mathbf{\Lambda}$.
Thus, the received signal-to-interference-and-noise ratio (SINR) of destination $i$ is given by
\begin{equation}
\operatorname{SINR}_i=\frac{P\left|(-1)^{i-1}(\gamma\sum_{m=1}^M|\det(\mathbf{H}_m)|)+[\mathbf{\Delta}\mathbf{\Gamma}\mathbf{H}]_{i,i}\right|^2}{1+\gamma\sum_{m=1}^M\left(|[\mathbf{H}_m]_{3-i,3-i}+[\mathbf{\Delta}_m]_{i,i}|^2+|[\mathbf{H}_m]_{i,3-i}+[\mathbf{\Delta}_m]_{i,3-i}|^2\right)+P|[\mathbf{\Delta}\mathbf{\Gamma}\mathbf{H}]_{i,3-i}|^2}.
\end{equation}
Define $R_i(\mathbf{Q})=\min_{\mathbf{A}\in\mathcal{A}_1(\mathbf{Q})}\log(1+\mbox{SINR}_i)$.
Then an achievable rate of destination $i$ is lower bounded by
\begin{equation}
R_i\geq\frac{1}{n_B}\sum_{\mathbf{Q}\in\mathcal{Q}_1}R_i(\mathbf{Q})\min\{\mbox{card}(\mathcal{T}_1(\mathbf{Q})),\mbox{card}(\mathcal{T}_2(F(\mathbf{Q})))\}.
\label{eq:r_i_b}
\end{equation}
From Lemmas \ref{lemma:prob} and \ref{lemma:large_num},
\begin{align}
\operatorname{card}(\mathcal{T}_1(\mathbf{Q}))\ge n_B(\P[\mathbf{H}[t]\in \mathcal{A}_1(\mathbf{Q})]-\delta)
\end{align}
and
\begin{align}
\operatorname{card}(\mathcal{T}_2(F(\mathbf{Q})))&\ge n_B(\P[\mathbf{G}[t]\in \mathcal{A}_2(F(\mathbf{Q}))]-\delta)\nonumber\\
&=n_B(\P[\mathbf{H}[t]\in \mathcal{A}_1(\mathbf{Q})]-\delta)
\end{align}
for all $\mathbf{Q}\in\mathcal{Q}_1$ with probability greater than $1-\frac{(2N+1)^{8M}}{n_B \delta^2}$, where we use $\operatorname{card}(\mathcal{Q}_1)=\operatorname{card}(\mathcal{Q}_2)=(2N+1)^{8M}$.
Then
\begin{align}
R_i&\ge\sum_{\mathbf{Q}\in\mathcal{Q}_1}R_i(\mathbf{Q})(\P[\mathbf{H}[t]\in \mathcal{A}_1(\mathbf{Q})]-\delta)\nonumber\\
&\ge\sum_{\mathbf{Q}\in\mathcal{Q}_1}R_i(\mathbf{Q})\P[\mathbf{H}[t]\in \mathcal{A}(\mathbf{Q})]-\delta2(2N+1)^{8M} \underset{\mathbf{Q}\in\mathcal{Q}_1}{\max}R_i(\mathbf{Q})
\end{align}
is achievable with probability greater than $1-\frac{(2N+1)^{8M}}{n_B \delta^2}$.
By setting $\Delta=n_B^{-1/(3\cdot 2^5M)}$, $N=n_B^{1/(3\cdot 2^4 M)}$, and $\delta=n_B^{-1/3}$, the following condition can be satisfied:
\begin{align}
\Delta&=n_B^{-1/(3\cdot 2^5M)}\to0,\nonumber\\
\Delta N&= n_b^{1/(3\cdot 2^5 M)}\to \infty,\nonumber\\
\delta2(2N+1)^{8M} \max_{\mathbf{Q}\in\mathcal{Q}_1}R_i(\mathbf{Q})&\leq 2\cdot 3^{8M}N^{8M}\delta\max_{\mathbf{Q}\in\mathcal{Q}_1}R_i(\mathbf{Q})\nonumber\\
&\overset{(a)}{\leq}2\cdot 3^{8M}N^{8M}\delta\log (1+2^4 M \Delta^2 N^2P)\nonumber\\
&=2\cdot 3^{8M}n_B^{-1/6}\log(1+2^4Mn_B^{1/(3\cdot 2^4M)}P)\to 0,\nonumber\\
\frac{(2N+1)^{8M}}{n_B\delta^2}&\leq \frac{3^{8M}N^{8M}}{n_B \delta^2}=3^{8M}n_B^{-1/6}\to 0
\end{align}
as $n_B$ increases, where $(a)$ follows since $|h_{ij}|^2\leq 2\Delta^2(N+\frac{1}{2})^2\leq2^3\Delta^2 N^2$ for the channel instances using the transmission (see Fig. \ref{fig:quan_1}).

Hence,
\begin{equation}
R_i=\E\left[\log \left(1+\frac{P\gamma^2\left(\sum_{m=1}^M|\det(\mathbf{H}_m)|\right)^2}{1+\gamma^2 \sum_{m=1}^M(|h_{2(m-1)+3-i,3-i}|^2+|h_{2(m-1)+i,3-i}|^2) }\right)\right]
\end{equation}
is achievable with probability approaching one  for $i\in\{1,2\}$, where we use the fact that
\begin{equation}
\lim_{\Delta\to 0}\mbox{SINR}_i=\frac{P\gamma^2\left(\sum_{m=1}^M|\det(\mathbf{H}_m)|\right)^2}{1+\gamma^2 \sum_{m=1}^M(|h_{2(m-1)+3-i,3-i}|^2+|h_{2(m-1)+i,3-i}|^2) }.
\end{equation}
In conclusion, Theorem \ref{th:sum_rate} holds.

\section{Approximate Capacity Characterization} \label{sec:appro_cap}
In this section, we prove Theorems \ref{th:constant_gap_unit_amp} to \ref{th:constant_gap_arb_L}, the approximate ergodic sum capacity characterization results.
We will deal with the difference between $R_{\operatorname{mimo}}$ and $R_{\operatorname{in}}$, which are given by \eqref{eq:R_in} and \eqref{eq:R_mimo} respectively.
Throughout this section, we assume a class of channel distributions such that $f(x)$ is only a function of $|x|$.
That is, for given amplitudes of the channel coefficients, their phases are i.i.d. uniformly distributed over $[0,2\pi)$.
For instance, this class of channel distributions includes i.i.d. uniform phase fading and i.i.d. Rayleigh fading as special cases.
We omit the time index $t$ in this section for notational convenience.

\subsection{Approximate Capacity for $L=2$} \label{subsec:l_2}

We first consider the case where $L=2$.
In order to deal with i.i.d. random phase in the rate expression in Theorem \ref{th:sum_rate}, we introduce the following lemma showing the exact solution of $\E_\phi\left[\log\left(1-x\cos\phi\right)\right]$ for $|x|\le1$ when  $\phi$ is uniformly distributed over $[0,2\pi)$.

\vspace{0.1in}
\begin{lemma} \label{lemma:uniform_phase}
Let $\phi$ be a random variable uniformly distributed over $[0,2\pi)$.
For $|x|\le1$,
\begin{equation}
\E_\phi\left[\log\left(1-x\cos\phi\right)\right]=\log\left(1+\sqrt{1-x^2}\right)-1.
\end{equation}
\end{lemma}
\vspace{0.1in}
\begin{proof}
We refer to the equation (4.224 12) in \cite{Jeffrey:07}.
\end{proof}
\vspace{0.1in}

\subsubsection{Proof of Theorem \ref{th:constant_gap_unit_amp}}
From \eqref{eq:R_in},
\begin{align} \label{eq:lower_unit_amp1}
R_{\operatorname{in}}  &\overset{(a)}{=} 2\E_{\theta}\left[\log\left(1+\frac{2P^2(1-\cos\theta)}{1+4P}\right)\right] \nonumber\\
&=2\log\left(1+\frac{2P^2}{1+4P}\right)+2\E_\theta\left[\log\left(1-\frac{2P^2}{1+4P+2P^2}\cos\theta\right)\right]\nonumber\\
&\overset{(b)}{=}2\log\left(1+\frac{2P^2}{1+4P}\right)+2\log\left(1+\sqrt{1-\left(\frac{2P^2}{1+4P+2P^2}\right)^2}\right)-2,
\end{align}
where $\theta=\theta_{1,1}+\theta_{2,2}-\theta_{1,2}-\theta_{2,1}$.
Here, $(a)$ follows since $|\det(\mathbf{H})|^2 =2(1-\cos\theta)$ and $\sigma_{\operatorname{AF},i}^2=\frac{2P}{1+2P}$,
$(b)$ follows since $\theta\!\!\mod\! [2\pi]$ is uniformly distributed over $[0,2\pi)$ and from  Lemma \ref{lemma:uniform_phase} with $|\frac{2P^2}{1+4P+2P^2}|\leq1$.
Similarly, from \eqref{eq:R_mimo},
\begin{align} \label{eq:upper_unit_amp1}
R_{\operatorname{mimo}}&= \E_\theta\left[\log((1+2P)^2-2P^2(1+\cos\theta))\right] \nonumber\\
&= \log(1+4P+2P^2)+\E_\theta\left[\log\left(1-\frac{2P^2}{1+4P+2P^2}\cos\theta\right)\right]\nonumber\\
&=  \log(1+4P+2P^2)+\log\left(1+\sqrt{1-\left(\frac{2P^2}{1+4P+2P^2}\right)^2}\right)-1.
\end{align}
Then, from \eqref{eq:lower_unit_amp1} and \eqref{eq:upper_unit_amp1},
\begin{align}
R_{\operatorname{mimo}}-R_{\operatorname{in}} &=\log\left(\frac{(1+4P)^2}{1+4P+2P^2}\right)-\log\left(1+\sqrt{1-\left(\frac{2P^2}{1+4P+2P^2}\right)^2}\right)+1\nonumber\\
&\overset{(a)}{\le}\log\left(\frac{(1+4P)^2}{1+4P+2P^2}\right)+1\nonumber\\
&\overset{(b)}{\le} 4,
\end{align}
where $(a)$ follows since $|\frac{2P^2}{1+4P+2P^2}|\le1$ for any $P>0$ and $(b)$ follows since
\begin{align}
\log\left(\frac{(1+4P)^2}{1+4P+2P^2}\right)&\le\log\left(\frac{(1+4P)^2}{1+2\sqrt{2}P+2P^2}\right)\nonumber\\
&=2\log\left(\frac{1+4P}{1+\sqrt{2}P}\right)\nonumber\\
&\leq 3,
\end{align}
where we use the fact that $\log\left(\frac{1+4P}{1+\sqrt{2}P}\right)$ is an increasing function of $P>0$ and $\lim_{P\to\infty}\log\left(\frac{1+4P}{1+\sqrt{2}P}\right)=\frac{3}{2}$.
In conclusion, Theorem \ref{th:constant_gap_unit_amp} holds.

\subsubsection{Proof of Theorem \ref{th:constant_gap_arb_amp_iid}}
Since $f(x)$ is only a function of $|x|$, $h_{i,j}$ can be represented as $a_{i,j}\exp(\jmath\theta_{i,j})$, where $a_{i,j}\ge 0$ and $\theta_{i,j}\in[0,2\pi)$ are independent of each other.
Moreover $\theta_{i,j}$ is uniformly distributed over $[0,2\pi)$.
To simplify the notation, we denote $\mathbf{a}=\{a_{1,1},a_{1,2},a_{2,1},a_{2,2}\}$, $A=a_{1,1}^2a_{2,2}^2+a_{1,2}^2a_{2,1}^2$, $B_1=a_{1,1}^2+a_{2,1}^2+2$, $B_2=a_{1,2}^2+a_{2,2}^2+2$, $G=2a_{1,1}a_{1,2}a_{2,1}a_{2,2}$, and $S = a_{1,1}^2+a_{1,2}^2+a_{2,1}^2+a_{2,2}^2$.

From \eqref{eq:R_in},
\begin{align} \label{eq:lower_niid1}
R_{\operatorname{in}}&\overset{(a)}{=}\sum_{i\in\{1,2\}}\E_{\mathbf{a}}\left[\E_{\theta}\left[\log \left(1+\frac{P^2(A-G\cos\theta)}{1+PB_i}\right)\right]\right],\nonumber\\
&=\sum_{i\in\{1,2\}}\E_{\mathbf{a}}\left[\log \left(1+\frac{P^2A}{1+PB_i}\right)\right]+\sum_{i\in\{1,2\}}\E_{\mathbf{a}}\left[\E_{\theta}\left[\log \left(1-\frac{P^2G\cos\theta}{1+PB_i+P^2 A}\right)\right]\right]\nonumber\\
&\overset{(b)}{=}\sum_{i\in\{1,2\}}\E_{\mathbf{a}}\left[\log \left(1+\frac{P^2A}{1+PB_i}\right)\right]+\sum_{i\in\{1,2\}}\E_{\mathbf{a}}\left[\log\left(1+\sqrt{1-\left(\frac{P^2G}{1+PB_i+P^2A}\right)^2}\right)\right]-2\nonumber\\
&\overset{(c)}{\ge}\sum_{i\in\{1,2\}}\E_{\mathbf{a}}\left[\log \left(1+\frac{P^2A}{1+PB_i}\right)\right]+2\E_{\mathbf{a}}\left[\log\left(1+\frac{\sqrt{A^2-G^2}}{A}\right)\right]-2\nonumber\\
&=\sum_{i\in\{1,2\}}\E_{\mathbf{a}}\left[\log \left(1+\frac{PA}{B_i}\right)\right]+\sum_{i\in\{1,2\}}\E_{\mathbf{a}}\left[\log \left(\frac{B_i+PB_i^2+P^2AB_i}{B_i+P(A+B_i^2)+P^2AB_i}\right)\right]\nonumber\\
&{~~~}+2\E_{\mathbf{a}}\left[\log\left(1+\frac{\sqrt{A^2-G^2}}{A}\right)\right]-2\nonumber\\
&\overset{(d)}{\ge}\sum_{i\in\{1,2\}}\E_{\mathbf{a}}\left[\log \left(1+\frac{PA}{B_i}\right)\right]+\sum_{i\in\{1,2\}}\E_{\mathbf{a}}\left[\log \left(\frac{B_i^2}{A+B_i^2}\right)\right]\nonumber\\
&{~~~}+2\E_{\mathbf{a}}\left[\log\left(1+\frac{\sqrt{A^2-G^2}}{A}\right)\right]-2,
\end{align}
where $\theta=\theta_{1,1}+\theta_{2,2}-\theta_{1,2}-\theta_{2,1}$.
Here $(a)$ follows from the facts that $\mathbf{a}$ and $\{\theta_{1,1},\theta_{1,2},\theta_{2,1},\theta_{2,2}\}$ are independent of each other and $|\det(\mathbf{H})|^2=A-G\cos\theta$, $(b)$ follows since $\theta\!\!\mod\! [2\pi]$ is uniformly distributed over $[0,2\pi)$ and from Lemma \ref{lemma:uniform_phase} with  $\left|\frac{P^2G}{1+PB_i+P^2A}\right|\leq1$, $(c)$ follows since $\frac{P^2G}{1+PB_i+P^2A}\le\frac{G}{A}$ for any $P\geq 0$, and $(d)$ follows  since $\log\left(\frac{c_1+c_2}{c_1+c_3}\right)\ge\log\left(\frac{c_2}{c_3}\right)$ for $c_1,c_2,c_3>0$ and  $c_2\le c_3$.

From \eqref{eq:R_mimo},
\begin{align} \label{eq:upper_iid3}
R_{\operatorname{mimo}}&\overset{(a)}{=}\E_{\mathbf{a}}\left[\E_{\theta}\left[\log\det(\mathbf{I}+P\mathbf{H}\mathbf{H}^\dagger)\right]\right] \nonumber\\
&=\E_{\mathbf{a}}\left[\E_{\theta}\left[\log\left(1+PS+P^2(A-G\cos\theta)\right)\right]\right]  \nonumber\\
&=\E_{\mathbf{a}}\left[\log(1+PS+P^2A)\right] + \E_{\mathbf{a}}\left[\E_{\theta}\left[\log\left(1-\frac{P^2G}{1+PS+P^2A}\cos\theta))\right)\right]\right]\nonumber\\
&\overset{(b)}{=}\E_{\mathbf{a}}\left[\log(1+PS+P^2A)\right] + \E_{\mathbf{a}}\left[\log\left(1+\sqrt{1-\left(\frac{P^2G}{1+PS+P^2A}\right)^2}\right)\right]-1\nonumber\\
&\overset{(c)}{\le}\E_{\mathbf{a}}\left[\log(1+PS+P^2A)\right],
\end{align}
where $(a)$ follows from the fact that $\mathbf{a}$ and $\{\theta_{1,1},\theta_{1,2},\theta_{2,1},\theta_{2,2}\}$ are independent of each other,
$(b)$ follows since $\theta\!\!\mod\! [2\pi]$ is uniformly distributed over $[0,2\pi)$ and from  Lemma \ref{lemma:uniform_phase} and $|\frac{P^2G}{1+PS+P^2A}|\leq1$, and $(c)$ follows again since $|\frac{P^2G}{1+PS+P^2A}|\leq1$.

Let
\begin{equation}
\Delta=\log(1+PS+P^2A)-\sum_{i\in\{1,2\}}\log \left(1+\frac{PA}{B_i}\right).
\end{equation}
Then
\begin{align} \label{eq:delta_iid}
\Delta&=\log\left(\frac{B_1B_2}{A}\right)+\log\left(\frac{1+PS+P^2 A}{\frac{B_1B_2}{A}+P(B_1+B_2)+P^2 A}\right)\nonumber\\
&\leq\log\left(\frac{B_1B_2}{A}\right),
\end{align}
where the inequality follows since $B_1B_2\geq A$ and $B_1+B_2\geq S$.
Therefore, from \eqref{eq:lower_niid1} to \eqref{eq:delta_iid},
\begin{align}
R_{\operatorname{mimo}}-R_{\operatorname{in}}&\leq\E_{\mathbf{a}}[\Delta]-\sum_{i\in\{1,2\}}\E_{\mathbf{a}}\left[\log \left(\frac{B_i^2}{A+B_i^2}\right)\right]-2\E_{\mathbf{a}}\left[\log\left(1+\frac{\sqrt{A^2-G^2}}{A}\right)\right]+2\nonumber\\
&\le\E_{\mathbf{a}}\left[\log\left(\frac{A(A+B_1^2)(A+B_2^2)}{B_1B_2(A+\sqrt{A^2-G^2})^2}\right)\right]+2\nonumber\\
&=2\E_{\mathbf{a}}\left[\log\left(\frac{\sqrt{A}(A+B_1^2)}{B_1(A+\sqrt{A^2-G^2})}\right)\right]+2.
\end{align}
In conclusion, Theorem \ref{th:constant_gap_arb_amp_iid} holds.

\subsection{Approximate Capacity as $L\to\infty$} \label{subsec:l_infty}

In this subsection, we characterize an approximate ergodic sum capacity in the limit of large number of relays by deriving $\lim_{L\to\infty}\{R_{\operatorname{mimo}}-R_{\operatorname{in}}\}$.
For $K$-user $2$-hop networks with $L$ relays, it was shown in \cite{Rankov:07} that interference can be completely neutralized if $K\geq N(N-1)+1$, which indicates that for $2\times L\times 2$ networks interference neutralization can be achieved without channel pairing if $L\geq3$.
However, maximizing the achievable sum rate exploiting interference neutralization without channel pairing presented in \cite{Esli:07} is non-convex and, as a result, it is unclear how to determine the sum rate gap from the cut-set upper bound.
By contrast, we now show that our achievable rate expression from Theorem \ref{th:sum_rate} permits to derive a finite-gap result.
The rate expression $R_{\operatorname{in}}$ in \eqref{eq:R_in} contains the sum of i.i.d. random variables, i.e., $\sum_{m=1}^M|\det(\mathbf{H}_m)|$, which approaches a deterministic value $M\E[|\det(\mathbf{H}_1)|]$ almost surely as $M\to\infty$ by the law of large numbers. The following lemma provides a rigorous lower bound in order to deal with $R_{\operatorname{in}}$ that holds for any $M$.

\vspace{0.10in}
\begin{lemma} \label{lemma:LLN}
Consider a sequence of i.i.d. nonnegative random variables $\{X_i, i\in\mathbb{Z}_+\}$.
Let $S_m=\sum_{i=1}^m X_i$.
If $\E[{X_1}^2]<\infty$, then for any $\epsilon\in(0,\E[X_1])$ and any $c>0$,
\begin{align}
\E\left[\log(1+c {S_m}^2)\right]\ge \log\left(1+cm^2(\E[X_1])^2\right)-\delta_m(c,\E[X_1],\E[{X_1}^2]),
\end{align}
where
\begin{align} \label{eq:def_delta}
\delta_m(c,\E[X_1],\E[{X_1}^2])&=\frac{\E[{X_1}^2]}{m\epsilon^2}\log\left(1+cm^2(\E[X_1]-\epsilon)^2\right)\nonumber\\
&{~~~}-\log\left(1-\frac{cm^2\epsilon(2\E[X_1]-\epsilon)}{1+cm^2(\E[X_1])^2}\right)
\end{align}
is a positive sequence of $m$, which converges to zero as $\epsilon\to 0$.
\end{lemma}
\vspace{0.10in}
\begin{proof}
We have
\begin{align}
&\E\left[\log(1+c {S_m}^2)\right]\nonumber \\
&=\E\left[\log\left(1+c{S_m}^2\right)\left(1_{\left\{|S_m/m-\E[X_1]|<\epsilon\right\}}+1_{\left\{|S_m/m-\E[X_1]|\ge\epsilon\right\}}\right)\right] \nonumber \\
&\overset{(a)}{\ge}\log\left(1+cm^2(\E[X_1])^2-2cm^2\epsilon\E[X_1]+cm^2\epsilon^2)\right)\E[1_{\left\{|S_m/m-\E[X_1]|<\epsilon\right\}}]\nonumber \\
&\overset{(b)}{\ge}\log\left(1+cm^2(\E[X_1])^2-2cm^2\epsilon\E[X_1]+cm^2\epsilon^2)\right)\left(1-\frac{\operatorname{Var}(X_1)}{m\epsilon^2}\right)\nonumber \\
&\overset{(c)}{\ge}\log\left(1+cm^2(\E[X_1])^2-2cm^2\epsilon\E[X_1]+cm^2\epsilon^2)\right)\nonumber\\
&{~~~}-\frac{\E[{X_1}^2]}{m\epsilon^2}\log\left(1+cm^2(\E[X_1])^2+cm^2\epsilon^2)\right)\nonumber \\
&=\log\left(1+cm^2(\E[X_1])^2\right)-\delta_m(c,\E[X_1],\E[{X_1}^2]),
\end{align}
where $(a)$ follows since $S_m>m \E[X_1]-m\epsilon$ under the condition $|S_m/m-\E[X_1]|<\epsilon$, $(b)$ follows from Chebyshev's inequality, and $(c)$ follows since $\operatorname{Var}(X_1)\le \E[{X_1}^2]$.
In conclusion, Lemma \ref{lemma:LLN} holds.
\end{proof}

By setting $\delta_m$ arbitrarily small as $m$ increases, Lemma \ref{lemma:LLN} provides that
\begin{equation}
\E\left[\log(1+c {S_m}^2)\right]\ge \log\left(1+cm^2(\E[X_1])^2\right)
\end{equation}
in the limit of large $m$.
Note that this bound is asymptotically tight since $\E\left[\log(1+c {S_m}^2)\right]\leq \log(1+c \E[{S_m}^2])$ from Jensen's inequality and $\log(1+c \E[{S_m}^2])$ is approximately given as $\log(1+cm^2\E[X_1]^2)$ as $m$ increases.

\subsubsection{Proof of Theorem \ref{th:constant_gap_unit_amp_L}}
Recall $M=\lfloor \frac{L}{2}\rfloor$. That is, $L\leq 2M+1$.
From \eqref{eq:R_mimo},
\begin{equation} \label{eq:jensen}
R_{\operatorname{mimo}}\le2\log(1+P(2M+1)),
\end{equation}
where the inequality follows from Jensen's inequality and the fact that $\log\det(\cdot)$ is a concave function \cite{Boyd:04}. Here we assume $L=2M+1$ to obtain an upper bound.

From  \eqref{eq:R_in},
\begin{align} \label{eq:r_in_lower}
R_{\operatorname{in}}&\overset{(a)}{=}2\E_{\{\theta_1,\cdots,\theta_M\}}\left[\log\left(1+\frac{P^2\left(\sum_{m=1}^M\sqrt{2-2\cos\theta_m}\right)^2}{1+P(2M+2)}\right)\right]\nonumber\\
&\overset{(b)}{\ge}2\log\left(1+\frac{\frac{16}{\pi^2}P^2M^2}{1+P(2M+2)}\right)-2\delta_M\left(\frac{P^2}{1+P(2M+2)},\frac{4}{\pi},2\right),
\end{align}
where $\theta_m=\theta_{2m-1,1}+\theta_{2m,2}-\theta_{2m-1,2}-\theta_{2m,1}$.
Here, $(a)$ follows since $|\det(\mathbf{H}_m)|=\sqrt{2-2\cos \theta_m}$ and $(b)$ follows since $\theta_m\!\!\mod\! [2\pi]$ is uniformly distributed over $[0,2\pi)$ and from Lemma \ref{lemma:LLN} with the facts that $\E\left[\sqrt{2-2\cos\theta_1}\right]=\frac{4}{\pi}$, and $\E\left[2-2\cos\theta_1\right]=2$.
Then, from \eqref{eq:jensen} and \eqref{eq:r_in_lower},
\begin{align}
R_{\operatorname{mimo}}-R_{\operatorname{in}}
&\leq 2\log\left(\frac{1+P(4M+3)+P^2(2M+1)(2M+2)}{1+P(2M+2)+\frac{16}{\pi^2}P^2M^2}\right)\nonumber\\
&{~~~}+2\delta_M\left(\frac{P^2}{1+P(2M+2)},\frac{4}{\pi},2\right).
\end{align}
Hence,
$\lim_{M\to\infty}\{R_{\operatorname{mimo}}-R_{\operatorname{in}}\}\leq 4\log \pi-4+\epsilon_1$,
where
\begin{align}
\epsilon_1&=\lim_{M\to\infty}2\delta_M\left(\frac{P^2}{1+P(2M+2)},\frac{4}{\pi},2\right)\nonumber\\
&=-2\log\left(1-\epsilon\left(\frac{\pi}{2}-\frac{\pi^2}{16}\epsilon\right)\right)>0,
\end{align}
which can be arbitrarily small as $\epsilon$ decreases.
In conclusion, Theorem \ref{th:constant_gap_unit_amp_L} holds.

\subsubsection{Proof of Theorem \ref{th:constant_gap_arb_L}}
From  \eqref{eq:R_in},
\begin{align} \label{r_in_2}
R_{\operatorname{in}}
&\ge 2\E\left[\log\left(1+P^2\left(\sum_{m=1}^M|\det(\mathbf{H}_m)|\right)^2\right)\right]\nonumber\\
&{~~~}-2\E\left[\log\left(1+P\left(\sum_{m=1}^{2M}|h_{m,1}|^2+2\right)\right)\right] \nonumber\\
&\overset{(a)}{\ge} 2\log\left(\frac{1+P^2M^2\left(\E[|\det(\mathbf{H}_1)|]\right)^2}{1+P(2M+2)}\right)\nonumber\\
&{~~~}-2\delta_M\left(P^2,\E[|\det(\mathbf{H}_1)|],\E\left[\left|\det(\mathbf{H}_1)\right|^2\right]\right),
\end{align}
where $(a)$ follows from Lemma \ref{lemma:LLN} and Jensen's inequality.
Hence, from \eqref{eq:jensen} and \eqref{r_in_2},
\begin{align}
R_{\operatorname{mimo}}-R_{\operatorname{in}}
&\leq 2\log\left(\frac{1+P(4M+3)+P^2(2M+1)(2M+2)}{1+P^2M^2\left(\E[|\det(\mathbf{H}_1)|]\right)^2}\right)\nonumber\\
&{~~~}+2\delta_M\left(P^2,\E[|\det(\mathbf{H}_1)|],\E\left[\left|\det(\mathbf{H}_1)\right|^2\right]\right)
\end{align}
and
\begin{align}
\lim_{M\to\infty}\left\{R_{\operatorname{mimo}}-R_{\operatorname{in}}\right\}\leq 4-4\log\left(\E[|\det(\mathbf{H}_1)|]\right)+\epsilon_2,
\end{align}
where
\begin{align}
\epsilon_2&=\lim_{M\to\infty}2\delta_M\left(P^2,\E[|\det(\mathbf{H}_1)|],\E\left[\left|\det(\mathbf{H}_1)\right|^2\right]\right)\nonumber\\
&=-2\log\left(1-\frac{\epsilon(2\E[|\det(\mathbf{H}_1)|]-\epsilon)}{(\E[|\det(\mathbf{H}_1)|])^2}\right)>0,
\end{align}
which can be arbitrarily small as $\epsilon$ decreases.
In conclusion, Theorem \ref{th:constant_gap_arb_L} holds.

\section{Conclusion} \label{sec:conclusion}
In this paper, we studied a fading $2$-user $2$-hop network with $L$ relays where channel coefficients vary over time.
In spite of recent achievements in this area, the best known capacity characterization is to within $o(\log \mbox{SNR})$ bits/sec/Hz from the ergodic sum capacity, which can be arbitrarily large as SNR increases. 
For a broad class of channel distributions, we tightened this gap to within a constant number of bits/sec/Hz, independent of SNR.
The achievability follows from ergodic interference neutralization in which the relays are partitioned into several pairs and interference is neutralized separately by each pair of relays.
The proposed scheme makes interference neutralized in the finite SNR regime and, at the same time, the intended signal power increased quadratically with $L$, leading that the optimal $2\log (L\mbox{SNR})$ rate scaling is achievable, which cannot be captured by the previous DoF work.

\section*{Appendix I\\Quantization for i.i.d. Uniform Phase Fading}

\begin{figure}[t!]
\begin{center}
\includegraphics[scale=0.9]{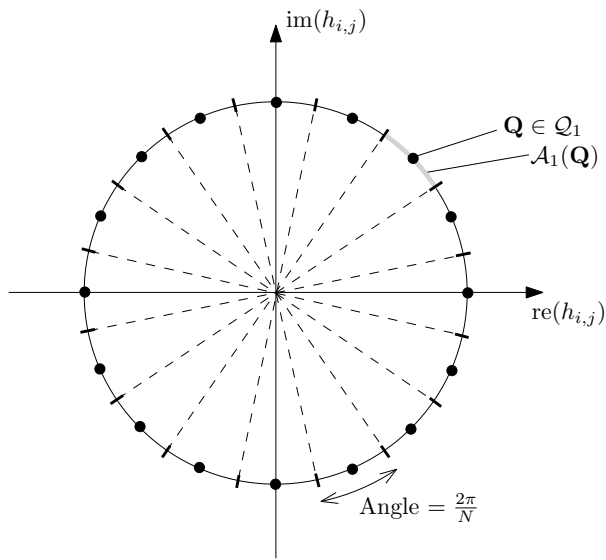}
\end{center}
\vspace{-0.15in}
\caption{Channel space partitioning with respect to the channel coefficient $h_{i,j}\in \mathcal{U}$ for i.i.d. uniform phase fading.}
\label{fig:quan_2}
\vspace{-0.1in}
\end{figure}

For i.i.d. uniform phase fading, $h_{i,j}[t]$ and $g_{j,i}[t]$ respectively are represented as $\exp(\jmath\theta_{i,j}[t])$ and $\exp(\jmath\varphi_{j,i}[t])$ for $i\in\{1,\cdots,L\}$ and $j\in\{1,2\}$.
Hence we can quantize the channel space of each hop based on angles.
Specifically, the channel space of the first hop can be partitioned as follows.
For $N\in\mathbb{Z}_+$, first define $\mathcal{Q}_1:=\big\{\exp(\jmath0),\exp(\jmath\frac{2\pi}{N}),\exp(\jmath\frac{4\pi}{N}),\cdots,\exp(\jmath\frac{(N-1)2\pi}{N})\big\}^{2M\times 2}$.
Let $\mathcal{U}$ denote the set of all $x\in\mathbb{C}$ satisfying $|x|=1$.
For a quantized channel matrix $\mathbf{Q}\in\mathcal{Q}_1$, define $\mathcal{A}_1(\mathbf{Q}):=\big\{\mathbf{A}\in\mathcal{U}^{2M\times 2}|-\frac{\pi}{N}\leq {\angle a_{i,j}}-\angle q_{i,j}<\frac{\pi}{N} \mbox{ for all } i\in\{1,\cdots,2M\} \mbox{ and }j\in\{1,2\}\big\}$, where $\mathbf{A}=\{a_{i,j}\}$, $\mathbf{Q}=\{q_{i,j}\}$, and $\angle{x}$ denotes the angle of  $x\in\mathcal{U}$, i.e., $x=\exp(\jmath \angle{x})$.
Figure \ref{fig:quan_2} illustrates the channel space partitioning with respect to $h_{i,j}\in \mathcal{U}$.
In a similar manner, we can define $\mathcal{Q}_2$ and $\mathcal{A}_2(\mathbf{Q})$ for the second hop.
Then we can show that there exists an increasing sequence of $N$, which is a function of $n_B$, such that \eqref{eq:achievable_rate} is achievable as $n_B$ increases using similar steps in the proof of Theorem \ref{th:sum_rate}.

\section*{Appendix II\\Closed Form of $\E[|\det(\mathbf{H}_1)|]$ for i.i.d. Rayleigh Fading}

Let $\mathbf{A}$ be a $2\times2$ matrix whose entries are i.i.d. circularly symmetric complex Gaussian random variables with mean zero and unit variance and $\mathbf{W}:= 2\mathbf{A}\mathbf{A}^\dagger$. Let $\lambda_1$ and $\lambda_2$,  $\lambda_1\ge \lambda_2$, be the eigenvalues of $\mathbf{W}$. Then the joint probability density function of $\lambda_1$ and $\lambda_2$ is given by \cite[Equation (3.11)]{edelman_thesis}
\begin{align}
f(\lambda_1,\lambda_2) = \frac{1}{16}\exp\left(-\frac{1}{2}(\lambda_1+\lambda_2)\right)(\lambda_1-\lambda_2)^2\mathbf{1}_{\{\lambda_1\ge\lambda_2\ge0\}}(\lambda_1,\lambda_2).
\end{align}
Thus,
\begin{align}
\E[|\det(\mathbf{H}_1)|] &= \frac{1}{2}\E\left[\sqrt{\det(\mathbf{W})}\right] \nonumber\\ 
&=\frac{1}{2}\E\left[\sqrt{\lambda_1\lambda_2}\right] \nonumber\\
&= \frac{1}{32} \int_0^\infty \int_0^{\lambda_1} \sqrt{\lambda_1\lambda_2}\exp\left(-\frac{1}{2}(\lambda_1+\lambda_2)\right)(\lambda_1-\lambda_2)^2 \, d\lambda_1 d\lambda_2 \nonumber\\
&\overset{(a)}{=} \frac{1}{32} \int_0^\infty \int_{0}^u \sqrt{(u+v)(u-v)}\exp\left(-u\right)(2v)^2 \,2 dv du \nonumber\\
&= \frac{1}{4} \int_0^\infty \left(\int_{0}^u v^2\sqrt{u^2-v^2}\, dv\right)\exp\left(-u\right) \, du \nonumber\\
&= \frac{1}{4} \int_0^\infty \frac{\pi}{16}u^4\exp(-u) du \nonumber\\
&= \frac{3\pi}{8},
\end{align}
where (a) follows by a change of variable $u=(\lambda_1+\lambda_2)/2$ and $v=(\lambda_1-\lambda_2)/2$.
In conclusion, $\E[|\det(\mathbf{H})|]=\frac{3\pi}{8}$ for i.i.d. Rayleigh fading.

\section*{Appendix III\\Approximate Ergodic Capacity for Fading Interference Channel}
A similar analysis used in Theorems \ref{th:constant_gap_unit_amp} and \ref{th:constant_gap_arb_amp_iid} is applicable for fading interference channel.
Specifically, consider the $K$-user interference channel in which the input--output relation is given by
\begin{equation} 
\mathbf{y}[t]=\mathbf{H}[t]\mathbf{x}[t]+\mathbf{z}[t]
\end{equation}
and the elements of $\mathbf{H}[t]=\{h_{i,j}[t]\}$ are i.i.d. drawn from a continuous distribution $f(x)$, $x\in\mathbb{C}$, and vary independently over time.
The achievability follows from ergodic interference alignment in \cite{Nazer11:09} showing that 
\begin{equation} \label{eq:lower_ic}
R_i=\frac{1}{2}\E[\log(1+2|h_{i,i}|^2P)]
\end{equation}
is achievable for all $i\in\{1,\cdots,K\}$ \cite[Theorem 2]{Nazer11:09}. 
Theorem \ref{th:eia} characterizes an approximate ergodic per-user capacity, i.e., ergodic sum capacity divided by $K$,  assuming that all sources employ uniform power allocation across time.
For this case, the sum of any pair of achievable rates is upper bounded by
\begin{align} \label{eq:upper_ic}
R_i+R_j \le \E\left[\log\left(1+\frac{(|h_{i,j}|^2+|h_{i,i}|^2)P}{\min\left\{1,\frac{|h_{i,j}|^2}{|h_{j,j}|^2}\right\}}\right)\right]
\end{align}
for all $i,j\in\{1,\cdots,K\}$, $i\neq j$ \cite[Equation (99)]{Nazer11:09}. 
From the lower bound \eqref{eq:lower_ic} and the upper bound \eqref{eq:upper_ic}, we characterize an approximate ergodic per-user capacity in the following theorem. 

\vspace{0.1in}
\begin{theorem} \label{th:eia}
Consider the fading $K$-user interference channel. Let $R_{\operatorname{ia}}:=\sum_{i=1}^K\frac{1}{2}\E[\log(1+2|h_{i,i}|^2P)]$ and $C_{\operatorname{sum}}$ denote the sum capacity assuming that all sources employ uniform power allocation across time.
Then
\begin{equation} \label{eq:gap_eia}
\frac{C_{\operatorname{sum}}-R_{\operatorname{ia}}}{K}\le \frac{1}{2}\log\left(\frac{3}{2}\right)+\frac{1}{2}\E\left[\left|\log\left(\frac{|h_{1,1}|^2}{|h_{1,2}|^2}\right)\right|\right]
\end{equation}
for any $P>0$.
\end{theorem}
\vspace{0.1in}
\begin{proof}
Define
\begin{align}
&\Delta(|h_{i,i}|^2,|h_{j,j}|^2,|h_{i,j}|^2) \nonumber \\
&:=\log\left(1+\frac{(|h_{i,j}|^2+|h_{i,i}|^2)P}{\min\left\{1,\frac{|h_{i,j}|^2}{|h_{j,j}|^2}\right\}}\right) - \frac{1}{2}\log(1+2|h_{i,i}|^2P) - \frac{1}{2}\log(1+2|h_{j,j}|^2P).
\end{align}
Then, from \eqref{eq:lower_ic} and \eqref{eq:upper_ic} and the fact that channel coefficients are i.i.d.,
\begin{equation} \label{eq:expected_gap}
\frac{C_{\operatorname{sum}}-R_{\operatorname{ia}}}{K}\le\frac{1}{2}\E[\Delta(|h_{1,1}|^2,|h_{2,2}|^2,|h_{1,2}|^2)].
\end{equation}
The term $\Delta(|h_{1,1}|^2,|h_{2,2}|^2,|h_{1,2}|^2)$ can be expressed as
\begin{align} \label{eq:delta}
&\Delta(|h_{1,1}|^2,|h_{2,2}|^2,|h_{1,2}|^2) \nonumber \\
=&\log\left(1+\max\left\{1,\frac{|h_{2,2}|^2}{|h_{1,2}|^2}\right\}(|h_{1,2}|^2+|h_{1,1}|^2)P\right) - \frac{1}{2}\log(1+2|h_{1,1}|^2P) - \frac{1}{2}\log(1+2|h_{2,2}|^2P) \nonumber \\
=&\frac{1}{2}\log\left(\frac{1+\max\left\{1,\frac{|h_{2,2}|^2}{|h_{1,2}|^2}\right\}(|h_{1,2}|^2+|h_{1,1}|^2)P}{1+2|h_{1,1}|^2P}\right) + \frac{1}{2}\log\left(\frac{1+\max\left\{1,\frac{|h_{2,2}|^2}{|h_{1,2}|^2}\right\}(|h_{1,2}|^2+|h_{1,1}|^2)P}{1+2|h_{2,2}|^2P}\right).
\end{align} 

The first term of~\eqref{eq:delta} is upper bounded as
\begin{align}
& \frac{1}{2}\log\left(\frac{1+\max\left\{1,\frac{|h_{2,2}|^2}{|h_{1,2}|^2}\right\}(|h_{1,2}|^2+|h_{1,1}|^2)P}{1+2|h_{1,1}|^2P}\right) \nonumber \\
&\le \frac{1}{2}\log\left(\frac{1+\max\left\{1,\frac{|h_{2,2}|^2}{|h_{1,2}|^2}\right\}|h_{1,2}|^2P+\max\left\{1,\frac{|h_{2,2}|^2}{|h_{1,2}|^2}\right\}2|h_{1,1}|^2P}{1+2|h_{1,1}|^2P}\right) \nonumber \\
&\le  \frac{1}{2}\log\left(\max\left\{1,\frac{|h_{2,2}|^2}{|h_{1,2}|^2}\right\}\frac{|h_{1,2}|^2}{2|h_{1,1}|^2}+\max\left\{1,\frac{|h_{2,2}|^2}{|h_{1,2}|^2}\right\}\right) \nonumber \\
&=\frac{1}{2}\log\left(\max\left\{1,\frac{|h_{2,2}|^2}{|h_{1,2}|^2}\right\}\right) +\frac{1}{2}\log\left(1+\frac{|h_{1,2}|^2}{2|h_{1,1}|^2}\right).
\end{align}
Similarly, the second term of~\eqref{eq:delta} is upper bounded as
\begin{align}
& \frac{1}{2}\log\left(\frac{1+\max\left\{1,\frac{|h_{2,2}|^2}{|h_{1,2}|^2}\right\}(|h_{1,2}|^2+|h_{1,1}|^2)P}{1+2|h_{2,2}|^2P}\right) \nonumber \\
&\le \frac{1}{2}\log\left(\frac{1+\max\left\{1,\frac{|h_{1,2}|^2}{|h_{2,2}|^2}\right\}2|h_{2,2}|^2P+\max\left\{1,\frac{|h_{2,2}|^2}{|h_{1,2}|^2}\right\}|h_{1,1}|^2P}{1+2|h_{2,2}|^2P}\right) \nonumber \\
&\le   \frac{1}{2}\log\left(\max\left\{1,\frac{|h_{1,2}|^2}{|h_{2,2}|^2}\right\}+\max\left\{1,\frac{|h_{2,2}|^2}{|h_{1,2}|^2}\right\}\frac{|h_{1,1}|^2}{2|h_{2,2}|^2}\right) \nonumber \\
&\overset{(a)}{=} \frac{1}{2}\log\left(\max\left\{1,\frac{|h_{1,2}|^2}{|h_{2,2}|^2}\right\}+\max\left\{1,\frac{|h_{1,2}|^2}{|h_{2,2}|^2}\right\}\frac{|h_{1,1}|^2}{2|h_{1,2}|^2}\right) \nonumber \\
&= \frac{1}{2}\log\left(\max\left\{1,\frac{|h_{1,2}|^2}{|h_{2,2}|^2}\right\}\right) +\frac{1}{2}\log\left(1+\frac{|h_{1,1}|^2}{2|h_{1,2}|^2}\right), 
\end{align}
where $(a)$ follows since $ a\max\{1,b/a\}=b\max\{1,a/b\}$ for all $a,b>0$.
Therefore, 
\begin{align} \label{eq:instant_gap}
&\Delta(|h_{1,1}|^2,|h_{2,2}|^2,|h_{1,2}|^2) \nonumber \\
&\le \frac{1}{2}\log\left(\max\left\{1,\frac{|h_{2,2}|^2}{|h_{1,2}|^2}\right\}\right) +\frac{1}{2}\log\left(1+\frac{|h_{1,2}|^2}{2|h_{1,1}|^2}\right)\nonumber\\
&{~~~} + \frac{1}{2}\log\left(\max\left\{1,\frac{|h_{1,2}|^2}{|h_{2,2}|^2}\right\}\right) +\frac{1}{2}\log\left(1+\frac{|h_{1,1}|^2}{2|h_{1,2}|^2}\right) \nonumber \\
&= \frac{1}{2} \left|\log\left(\frac{|h_{2,2}|^2}{|h_{1,2}|^2}\right)\right| +\frac{1}{2}\log\left(1+\frac{|h_{1,2}|^2}{2|h_{1,1}|^2}\right) +\frac{1}{2}\log\left(1+\frac{|h_{1,1}|^2}{2|h_{1,2}|^2}\right) \nonumber \\
&\le  \frac{1}{2} \left|\log\left(\frac{|h_{2,2}|^2}{|h_{1,2}|^2}\right)\right| +\frac{1}{2}\log\left(1+\frac{1}{2}\right) +\frac{1}{2}\log\left(1+\frac{1}{2}\max\left\{\frac{|h_{1,2}|^2}{|h_{1,1}|^2},\frac{|h_{1,1}|^2}{|h_{1,2}|^2}\right\}\right) \nonumber \\
&\le \frac{1}{2}\log\left(\frac{3}{2}\right) +\frac{1}{2} \left|\log\left(\frac{|h_{2,2}|^2}{|h_{1,2}|^2}\right)\right| + \frac{1}{2}\log\left(\frac{3}{2}\max\left\{\frac{|h_{1,2}|^2}{|h_{1,1}|^2},\frac{|h_{1,1}|^2}{|h_{1,2}|^2}\right\}\right) \nonumber \\
&= \log\left(\frac{3}{2}\right) +\frac{1}{2} \left|\log\left(\frac{|h_{2,2}|^2}{|h_{1,2}|^2}\right)\right| + \frac{1}{2}\left|\log\left(\frac{|h_{1,1}|^2}{|h_{1,2}|^2}\right)\right|.
\end{align}
Finally, combining \eqref{eq:expected_gap} and \eqref{eq:instant_gap} shows the gap in \eqref{eq:gap_eia}, which completes the proof.
\end{proof}

For i.i.d. Rayleigh fading channel,~$|h_{i,j}|^2$ has the exponential distribution and
\begin{align}
f_{|h_{1,1}|^2/|h_{1,2}|^2}(x) = \frac{1}{(x+1)^2}
\end{align}
for $x\ge 0$.
Therefore,
\begin{align}
\E\left[\left|\log\left(\frac{|h_{1,1}|^2}{|h_{1,2}|^2}\right)\right|\right] \nonumber & = \int_{0}^\infty\frac{|\log x|}{(x+1)^2}\,  dx = 2
\end{align}
and the gap in Theorem \ref{th:eia} is given as $\frac{1}{2}\log 6$ bits/sec/Hz (approximately $1.3$ bits/sec/Hz).


\begin{thebibliography}{10}
\providecommand{\url}[1]{#1}
\csname url@samestyle\endcsname
\providecommand{\newblock}{\relax}
\providecommand{\bibinfo}[2]{#2}
\providecommand{\BIBentrySTDinterwordspacing}{\spaceskip=0pt\relax}
\providecommand{\BIBentryALTinterwordstretchfactor}{4}
\providecommand{\BIBentryALTinterwordspacing}{\spaceskip=\fontdimen2\font plus
\BIBentryALTinterwordstretchfactor\fontdimen3\font minus
  \fontdimen4\font\relax}
\providecommand{\BIBforeignlanguage}[2]{{%
\expandafter\ifx\csname l@#1\endcsname\relax
\typeout{** WARNING: IEEEtran.bst: No hyphenation pattern has been}%
\typeout{** loaded for the language `#1'. Using the pattern for}%
\typeout{** the default language instead.}%
\else
\language=\csname l@#1\endcsname
\fi
#2}}
\providecommand{\BIBdecl}{\relax}
\BIBdecl

\bibitem{Etkin:08}
R.~H. Etkin, D.~N.~C. Tse, and H.~Wang, ``Gaussian interference channel
  capacity to within one bit,'' \emph{{IEEE} Trans. Inf. Theory}, vol.~54, pp.
  5534--5562, Dec. 2008.

\bibitem{Viveck1:08}
V.~R. Cadambe and S.~A. Jafar, ``Interference alignment and degrees of freedom
  of the {$K$}-user interference channel,'' \emph{{IEEE} Trans. Inf. Theory},
  vol.~54, pp. 3425--3441, Aug. 2008.

\bibitem{Nazer11:09}
B.~Nazer, M.~Gastpar, S.~A. Jafar, and S.~Vishwanath, ``Ergodic interference
  alignment,'' \emph{{IEEE} Trans. Inf. Theory}, vol.~58, pp. 6355--6371, Oct.
  2012.

\bibitem{Mohajer:11}
S.~Mohajer, S.~Diggavi, C.~Fragouli, and D.~N.~C. Tse, ``Approximate capacity
  of a class of {G}aussian interference-relay networks,'' \emph{{IEEE} Trans.
  Inf. Theory}, vol.~57, pp. 2837--2864, May 2011.

\bibitem{Tiangao:12}
T.~Gou, S.~A. Jafar, C.~Wang, S.-W. Jeon, and S.-Y. Chung, ``Aligned
  interference neutralization and the degrees of freedom of the
  $2\times2\times2$ interference channel,'' \emph{{IEEE} Trans. Inf. Theory},
  vol.~58, pp. 4381--4395, Jul. 2012.

\bibitem{Jeon2:11}
S.-W. Jeon, S.-Y. Chung, and S.~A. Jafar, ``Degrees of freedom region of a
  class of multisource {G}aussian relay networks,'' \emph{{IEEE} Trans. Inf.
  Theory}, vol.~57, pp. 3032--3044, May 2011.

\bibitem{Wang:11}
C.~Wang, T.~Gou, and S.~A. Jafar, ``Multiple unicast capacity of $2$-source
  $2$-sink networks,'' in \emph{arXiv:cs.IT/1104.0954}, Apr. 2011.

\bibitem{Shomorony:11}
I.~Shomorony and A.~S. Avestimehr, ``Two-unicast wireless networks:
  {C}haracterizing the degrees-of-freedom,'' in \emph{arXiv:cs.IT/1102.2498},
  Feb. 2011.

\bibitem{Rankov:07}
B.~Rankov and A.~Wittneben, ``Spectral efficient protocols for half-duplex
  fading relay channels,'' \emph{{IEEE} J. Sel. Areas Commun.}, vol.~25, pp.
  379--389, Feb. 2007.

\bibitem{Viveck2:09}
V.~R. Cadambe and S.~A. Jafar, ``Interference alignment and the degrees of
  freedom of wireless {$X$} networks,'' \emph{{IEEE} Trans. Inf. Theory},
  vol.~55, pp. 3893--3908, Sep. 2009.

\bibitem{Nazer:09}
B.~Nazer, M.~Gastpar, S.~A. Jafar, and S.~Vishwanath, ``Ergodic interference
  alignment,'' in \emph{Proc. {IEEE} Int. Symp. Information Theory (ISIT)},
  Seoul, South Korea, Jun./Jul. 2009.

\bibitem{Maddah-Ali:08}
M.~A. Maddah-Ali, A.~S. Motahari, and A.~K. Khandani, ``Communication over
  {MIMO} {X} channels: {I}nterference alignment, decomposition, and performance
  analysis,'' \emph{{IEEE} Trans. Inf. Theory}, vol.~54, pp. 3457--3470, Aug.
  2008.

\bibitem{Tiangao:10}
T.~Gou and S.~A. Jafar, ``Degrees of freedom of the {$K$} user {$M\times N$}
  {MIMO} interference channel,'' \emph{{IEEE} Trans. Inf. Theory}, vol.~56, pp.
  6040--6057, Dec. 2010.

\bibitem{Suh:11}
C.~Suh, M.~Ho, and D.~N.~C. Tse, ``Downlink interference alignment,''
  \emph{{IEEE} Trans. Commun.}, vol.~59, pp. 2616--2626, Sep. 2011.

\bibitem{Suh:08}
C.~Suh and D.~N.~C. Tse, ``Interference alignment for cellular networks,'' in
  \emph{Proc. 46th Annu. Allerton Conf. Communication, Control, and Computing},
  Monticello, IL, Sep. 2008.

\bibitem{Viveck1:09}
V.~R. Cadambe and S.~A. Jafar, ``Degrees of freedom of wireless networks with
  relays, feedback, cooperation, and full duplex operation,'' \emph{{IEEE}
  Trans. Inf. Theory}, vol.~55, pp. 2334--2344, May 2009.

\bibitem{Annapureddy:11}
V.~S. Annapureddy, A.~{El Gamal}, and V.~V. Veeravalli, ``Degrees of freedom of
  interference channels with {CoMP} transmission and reception,'' \emph{{IEEE}
  Trans. Inf. Theory}, vol.~58, pp. 5740--5760, Sep. 2012.

\bibitem{Ke:12}
L.~Ke, A.~Ramamoorthy, Z.~Wang, and H.~Yin, ``Degrees of freedom region for an
  interference network with general message demands,'' \emph{{IEEE} Trans. Inf.
  Theory}, vol.~58, pp. 3787--3797, Jun. 2012.

\bibitem{Motahari:09}
A.~S. Motahari, S.~O. Gharan, and A.~K. Khandani, ``Real interference alignment
  with real numbers,'' in \emph{arXiv:cs.IT/0908.1208}, 2009.

\bibitem{Motahari2:09}
A.~S. Motahari, S.~O. Gharan, M.~A. Maddah-Ali, and A.~K. Khandani, ``Real
  interference alignment: {E}xploiting the potential of single antenna
  systems,'' in \emph{arXiv:cs.IT/0908.2282}, 2009.

\bibitem{Bresler:10}
G.~Bresler, A.~Parekh, and D.~N.~C. Tse, ``The approximate capacity of the
  many-to-one and one-to-many {G}aussian interference channels,'' \emph{{IEEE}
  Trans. Inf. Theory}, vol.~56, pp. 4566--4592, Sep. 2010.

\bibitem{Nam:10}
W.~Nam, S.-Y. Chung, and Y.~H. Lee, ``Capacity of the {G}aussian two-way relay
  channel to within $\frac{1}{2}$ bit,'' \emph{{IEEE} Trans. Inf. Theory},
  vol.~56, pp. 5488--5494, Nov. 2010.

\bibitem{Avestimehr:11}
A.~S. Avestimehr, S.~N. Diggavi, and D.~N.~C. Tse, ``Wireless network
  information flow: {A} deterministic approach,'' \emph{{IEEE} Trans. Inf.
  Theory}, vol.~57, pp. 1872--1905, Apr. 2011.

\bibitem{Lim:11}
S.~H. Lim, Y.-H. Kim, A.~{El Gamal}, and S.-Y. Chung, ``Noisy network coding,''
  \emph{{IEEE} Trans. Inf. Theory}, vol.~57, pp. 3132 -- 3152, May 2011.

\bibitem{Niesen2:11}
U.~Niesen and M.~A. Maddah-Ali, ``Interference alignment: {F}rom
  degrees-of-freedom to constant-gap capacity approximations,'' in
  \emph{arXiv:cs.IT/1112.4879}, 2011.

\bibitem{Ordentlich:12}
O.~Ordentlich, U.~Erez, and B.~Nazer, ``The approximate sum capacity of the
  symmetric {G}aussian $k$-user interference channel,'' in
  \emph{arXiv:cs.IT/1206.0197}, 2012.

\bibitem{JeonITA:09}
S.-W. Jeon and S.-Y. Chung, ``Capacity of a class of multi-source relay
  networks,'' in \emph{Information Theory and Applications Workshop},
  University of California San Diego, La Jolla , CA, Feb. 2009.

\bibitem{Niesen:11}
U.~Niesen, B.~Nazer, and P.~Whiting, ``Computation alignment: {C}apacity
  approximation without noise accumulation,'' in \emph{arXiv:cs.IT/1108.6312},
  2011.

\bibitem{Telatar:99}
I.~E. Telatar, ``Capacity of multi-antenna {G}aussian channels,''
  \emph{European Trans. on Telecommun.}, vol.~10, pp. 585--595, Nov. 1999.

\bibitem{Csiszar:81}
I.~Csisz{\'a}r and J.~K{\"o}rner, \emph{Information Theory: Coding Theorems for
  Discrete Memoryless Systems}.\hskip 1em plus 0.5em minus 0.4em\relax New
  York: Academic Press, 1981.

\bibitem{Jeffrey:07}
I.~S. Gradshteyn and I.~M. Ryzhik, \emph{Table of Integrals, Series and
  Products}.\hskip 1em plus 0.5em minus 0.4em\relax Academic Press, 2007.

\bibitem{Esli:07}
C.~Esli, S.~Berger, and A.~Wittneben, ``Optimizing zero-forcing based gain
  allocation for wireless multiuser networks,'' in \emph{Proc. {IEEE}
  {International Conference on Communications (ICC)}}, Beijing, China, Jun.
  2007.

\bibitem{Boyd:04}
S.~Boyd and L.~Vandenberghe, \emph{Convex Optimization}.\hskip 1em plus 0.5em
  minus 0.4em\relax New York: Cambridge Univ. Press, 2004.

\bibitem{edelman_thesis}
A.~Edelman, ``Eigenvalues and condition numbers of random matrices,'' Ph.D.
  dissertation, Massachusetts Institute of Technology (MIT), 1989.

\end{thebibliography}


\end{document}